\def\isarxiv{1}

\ifdefined\isarxiv
\documentclass[a4paper, twocolumn, 10pt]{article}
\usepackage[top=2cm, bottom=2cm, left=1.75cm, right=1.75cm]{geometry}
\usepackage[hyphens]{url}
\usepackage{hyperref}
\date{}
\else
\documentclass[conference]{IEEEtran}
\fi

\usepackage[T1]{fontenc}
\usepackage{subcaption}

\usepackage{pifont}
\usepackage{algorithmic}
\usepackage{textcomp}
\usepackage[dvipsnames]{xcolor}
\usepackage{paralist}
\usepackage{enumitem}
\usepackage{hyperref}

\usepackage{float}
\usepackage{fvextra}
\usepackage[textsize=footnotesize]{todonotes}
\usepackage{glossaries}
\glsdisablehyper
\usepackage{caption}
\usepackage{listings}
\usepackage{xspace}
\usepackage{comment}
\usepackage{fancyvrb}
\usepackage{graphicx,txfonts}
\usepackage{multirow}
\usepackage{tikz}
\usepackage{multicol}
\usepackage{verbatim}
\usepackage{framed}
\usepackage{setspace}
\usepackage{fontenc}
\usepackage{bytefield}
\usepackage{booktabs}
\usepackage[dvipsnames]{xcolor}
\usepackage{mwe}
\usepackage{wrapfig}
\usepackage{color, colortbl}
\usepackage{threeparttable}
\usepackage{array}
\usepackage{adjustbox}
\usepackage{wasysym}
\usepackage{xparse}
\usepackage[htt]{hyphenat}
\usepackage[hashEnumerators,smartEllipses]{markdown}
\usepackage{todonotes}
\usepackage{censor}
\usepackage{cleveref}

\usepackage{tikz}
\usetikzlibrary{arrows.meta,positioning,shapes.geometric,fit}
\newif\ifabridged
\newif\ifnotabridged
\newif\ifanonymous
\newif\ifnotanonymous
\newif\ifreviewer
\newif\ifnotreviewer
\newif\iftags
\newif\ifnottags

\ifdefined\isabridged
\abridgedtrue
\fi

\ifabridged\notabridgedfalse
\else\notabridgedtrue
\fi

\ifdefined\isanonymous
\anonymoustrue
\fi

\ifanonymous\notanonymousfalse
\else\notanonymoustrue
\fi

\newif\ifarxiv
\newif\ifnotarxiv

\ifdefined\isarxiv
\arxivtrue
\fi

\newcounter{tncnt}
\newcounter{mgcnt}
\newcounter{accnt}

\newcounter{hecnt}

\Crefname{section}{\S$\!$}{\S\S$\!$}
\Crefname{figure}{Fig.}{Figs.}
\crefname{figure}{Fig.}{Figs.}

\renewcommand{\paragraph}{\noindent\textbf}

\newcommand{\dCOne}{\ding{192}\xspace}
\newcommand{\dCTwo}{\ding{193}\xspace}
\newcommand{\dCThree}{\ding{194}\xspace}
\newcommand{\dCFour}{\ding{195}\xspace}
\newcommand{\dCFive}{\ding{196}\xspace}

\newcommand{\prv}{{\ensuremath{\sf{Prv}}}\xspace}
\newcommand{\vrf}{{\ensuremath{\sf{Vrf}}}\xspace}

\renewcommand{\paragraph}{\noindent\textbf}

\newcommand{\acron}{\texttt{PRISM}\xspace}
\newcommand{\PRISM}{\acron}
\newcommand{\enclavecap}{prismatic capability\xspace}
\newcommand{\Enclavecap}{Prismatic capability\xspace}
\newcommand{\enclavecaps}{prismatic capabilities\xspace}
\newcommand{\Enclavecaps}{Prismatic capabilities\xspace}
\newcommand{\capcolor}{hue\xspace}
\newcommand{\Capcolor}{Hue\xspace}
\newcommand{\capcolors}{hues\xspace}
\newcommand{\Capcolors}{Hues\xspace}
\newcommand{\gradient}{color\xspace}
\newcommand{\gradients}{colors\xspace}

\newcommand{\capvalidtable}{\gls{cam}\xspace}
\newcommand{\caphashtable}{hue table\xspace}
\newcommand{\tcb}{\gls{hmm}\xspace}

\newcommand{\EInvoke}{\texttt{EInvoke}\xspace}
\newcommand{\EExit}{\texttt{EExit}\xspace}
\newcommand{\PICASSO}{\acrshort{picasso}\xspace}

\newcommand{\ccs}{\glsentrylongpl{cc}\xspace}

\newcommand{\CCs}{\Glsentrylongpl{cc}\xspace}
\newcommand{\cchyph}{\glsentryhyph{cc}\xspace}

\newcommand{\ccpid}{\gls{pid}\xspace}

\newcommand{\ccactive}{valid\xspace}
\newcommand{\ccinactive}{invalid\xspace}
\usepackage{tikz}
\usetikzlibrary{arrows}
\usetikzlibrary{shapes}
\newcommand*\circled[3]{\tikz[baseline=(char.base)]{
            \node[scale=.7,shape=circle,draw,inner sep=1pt,fill=#2,minimum size=.1cm,text=#3] (char) {#1};}}

\newcommand{\colorbitbox}[3]{
  \sbox0{\bitbox{#2}{#3}}
  \makebox[0pt][l]{\textcolor{#1}{\rule[-\dp0]{\wd0}{\ht0}}}
  \bitbox{#2}{#3}
}

\newcommand{\cmark}{\textcolor{green!55!black}{\ding{51}}}
\newcommand{\xmark}{\textcolor{red!70!black}{\ding{55}}}

\newcommand{\appSubPlotWidth}{0.095\textwidth}
\newcommand{\appSubPlotHeight}{2cm}
\newcommand{\appSubPlotHeightN}{1.5cm}

\makeatletter
\newcommand{\myfnsymbol}[1]{%
  \expandafter\@myfnsymbol\csname c@#1\endcsname
}
\newcommand{\@myfnsymbol}[1]{%
  \ifcase #1
  \or \TextOrMath{\textasteriskcentered}{*}%
  \or \TextOrMath{\textdagger}{\dagger}%
  \or \TextOrMath{$\ddagger$}{\ddagger}%
  \or \TextOrMath{\textsection}{\S}%
  \or \TextOrMath{\textbardbl}{\|}%
  \or \TextOrMath{\P}{\P}%
  \or \TextOrMath{\#}{\#}%
  \or \TextOrMath{$\Delta$}{\Delta}%
  \fi
}
\newcommand{\equalContribution}{\@myfnsymbol{1}}
\newcommand{\affiliationA}{\@myfnsymbol{2}}
\newcommand{\affiliationB}{\@myfnsymbol{3}}
\newcommand{\affiliationC}{\@myfnsymbol{4}}
\newcommand{\affiliationD}{\@myfnsymbol{5}}
\newcommand{\affiliationE}{\@myfnsymbol{6}}
\newcommand{\affiliationF}{\@myfnsymbol{7}}
\newcommand{\affiliationG}{\@myfnsymbol{8}}
\makeatother

\glsaddkey
    {hyphenated}
    {\relax}
    {\glsentryhyph}
    {\Glsentryhyph}
    {\glshyph}
    {\Glshyph}
    {\GLShyph}

\DeclareFontFamily{OT1}{mathc}{}
\DeclareFontShape{OT1}{mathc}{m}{it}{<-> mathc10}{}
\DeclareMathAlphabet{\mathabxcal}{OT1}{mathc}{m}{it}
\newacronym{abi}{ABI}{application binary interface}
\newacronym{alloc}{$\mathabxcal{Alloc}$}{allocator}
\newacronym{alu}{ALU}{arithmetic logic unit}
\newacronym{api}{API}{application programming interface}
\newacronym{dma}{DMA}{direct memory access}
\newacronym{caprelocs}{\texttt{\_\!\_cap\_relocs}}{capability relocation table}
\newacronym[hyphenated={colored-capability},longplural={colored capabilities}]{cc}{CC}{colored capability}
\newacronym{cisa}{CISA}{Cybersecurity & Infrastructure Security Agency}
\newacronym{cca}{CCA}{Confidential Computing Architecture}
\newacronym{cheri}{CHERI}{Capability Hardware Enhanced RISC Instructions}
\newacronym{clc}{\texttt{clc}}{load capability via capability}
\newacronym{cove}{CoVE}{Confidential VM Extension}
\newacronym{cpu}{CPU}{central processing unit}
\newacronym{crg}{\texttt{CRG}}{capability read generation }
\newacronym{csc}{\texttt{csc}}{store capability via capability}
\newacronym{csp}{\texttt{csp}}{stack pointer capability}
\newacronym{csr}{CSR}{control and status register}
\newacronym{cve}{CVE}{common vulnerability enumeration}
\newacronym{cvm}{CVM}{confidential virtual machine}
\newacronym{cwe}{CWE}{common weakness enumeration}
\newacronym{cw}{\texttt{CW}}{capability write}
\newacronym{ddc}{\texttt{ddc}}{default data capability}
\newacronym{dcc}{\texttt{ddc}}{default code capability}
\newacronym{df}{DF}{double-free}
\newacronym{eda}{EDA}{electronic design automation}
\newacronym{edmm}{EDMM}{Enclave Dynamic Memory Management}
\newacronym{epc}{EPC}{Enclave Page Cache}
\newacronym{epvt}{EPVT}{enclave provenance-validity table}
\newacronym{edl}{EDL}{Enclave Definition Language}
\newacronym{ffi}{FFI}{foreign function interface}
\newacronym{fpga}{FPGA}{field-programmable gate array}

\newacronym{hmm}{HMM}{\Capcolor Management Monitor}
\newacronym{id}{ID}{identifier}
\newacronym[longplural={instruction set architectures}]{isa}{ISA}{instruction-set architecture}
\newacronym{ir}{IR}{intermediate representation}
\newacronym{ip}{IP}{intellectual property}
\newacronym{ipi}{IPI}{inter-processor interrupt}
\newacronym{jit}{JIT}{just-in-time}
\newacronym{lsu}{LSU}{load-store unit}
\newacronym{lsq}{LSQ}{Load/Store Queue}
\newacronym{gep}{\textsf{GEP}}{\textsf{GetElementPtr}}
\newacronym[hyphenated={provenance-identifier},longplural={provenance identifiers}]{pid}{provenance ID}{provenance identifier}
\newacronym{ncsc}{NCSC}{National Cyber Security Centre}
\newacronym{nist}{NIST}{National Institute of Standards and Technology}
\newacronym{mte}{MTE}{Memory Tagging Extension}
\newacronym{mtt}{MTT}{Memory Tracking Table}
\newacronym{mrs}{MRS}{\texttt{malloc} revocation shim}
\newacronym{mmu}{MMU}{memory management unit}
\newacronym{m-mode}{M-mode}{machine mode}
\newacronym{mpx}{MPX}{Memory Protection Extension}
\newacronym{napot}{NAPOT}{naturally aligned power-of-two}
\newacronym[hyphenated={operating-system}]{os}{OS}{operating system}
\newacronym{otype}{\texttt{otype}}{object type}
\newacronym{otth}{OTYPETH}{otype threshold}
\newacronym{u}{\texttt{U}}{user mode access allowed}
\newacronym{ucrg}{\texttt{UCRG}}{user capability read generation}
\newacronym{pa}{PA}{pointer authentication}
\newacronym{pc}{\texttt{pc}}{program counter}
\newacronym{pcc}{\texttt{pcc}}{program counter capability}
\newacronym{picasso}{\texttt{PICASSO}}{Provenance Indirection-enabled CHERI Architecture for Scarcely Swept Objects}
\newacronym{poc}{PoC}{proof-of-concept}
\newacronym[hyphenated={pure-capability}]{purecap}{purecap}{pure capability}
\newacronym{pte}{PTE}{page table entry}
\newacronym{ptlb}{PTLB}{provenance-translation lookaside buffer}
\newacronym{pvb}{PVB}{provenance-validity bit}
\newacronym{pvt}{PVT}{provenance-validity table}
\newacronym{pvtr}{PVTR}{provenance-validity table register}
\newacronym{ross}{\texttt{ROSS}}{Revocation-Orchestrating System Service}
\newacronym{rtl}{RTL}{register-transfer level}
\newacronym{rtos}{RTOS}{real-time operating system}
\newacronym{rss}{RSS}{resident set size}
\newacronym{rv64y}{RV64Y}{64-bit RISC-V}
\newacronym{rof}{RoF}{revoke-on-free}

\newacronym{sard}{SARD}{Software Assurance Reference Dataset}
\newacronym{sbi}{SBI}{Supervisor Binary Interface}
\newacronym{sdk}{SDK}{software development kit}
\newacronym{sgx}{SGX}{Software Guard Extension}
\newacronym{ssa}{SSA}{state save area}
\newacronym{sp}{\texttt{sp}}{stack pointer}
\newacronym{soc}{SoC}{system-on-chip}
\newacronym{tdx}{TDX}{Trust Domain Extension}
\newacronym{tlb}{TLB}{translation lookaside buffer}
\newacronym{toctou}{TOCTOU}{time-of-check to time-of-use}
\newacronym{tps}{TPS}{transactions per second}
\newacronym{trts}{TRTS}{trusted enclave runtime system}
\newacronym{tsm}{TSM}{TEE Security Manager}
\newacronym{tvm}{TVM}{TEE Virtual Machine}
\newacronym{urts}{URTS}{untrusted runtime system}

\newacronym{uaf}{UAF}{use-after-free}
\newacronym{uar}{UAR}{use-after-reallocation}
\newacronym{unr}{\texttt{unr}}{unit number}
\newacronym{vm}{VM}{virtual machine}
\newacronym{wns}{WNS}{worst negative slack}
\newacronym{qps}{QPS}{Queries Per Second}
\newacronym{tcb}{TCB}{trusted computing base}
\newcommand{\adv}{\ensuremath{\sf{\mathcal Adv}}\xspace}
\newacronym{adv}{\adv}{adversary}
\newacronym{cam}{HAM}{Hue-Addressed Memory}

\newacronym{tee}{TEE}{trusted execution environment}
\newacronym{pmp}{PMP}{physical memory protection}
\newacronym{ocall}{OCALL}{outside call}
\newacronym{ecall}{ECALL}{enclave call}

\makeatletter
\newcommand{\crefnames}[3]{%
  \@for\next:=#1\do{%
    \expandafter\crefname\expandafter{\next}{#2}{#3}%
  }%
}
\makeatother

\crefnames{part,chapter,section}{\S}{\S\S}
\usepackage[utf8]{inputenc}
\usepackage{pgfplots, pgfplotstable}
\pgfplotsset{compat=1.18}
\usepgfplotslibrary{groupplots}
\definecolor{barwhite}{gray}{1.0}
\definecolor{barlightgray}{gray}{0.75}
\definecolor{bardarkgray}{gray}{0.40}
\definecolor{barblack}{gray}{0.0}

\usepackage{listings}
\lstdefinelanguage{Tamarin}{
  keywords={rule, let, in, builtins, equations, restriction, lemma,
            all-traces, exists-trace, Fr, In, Out, theory, begin, end, pk, All, Ex, not, h, K, reuse},
  keywordstyle=\color{blue}\bfseries,
  morecomment=[l]{//},
  morecomment=[l]{///},
  morecomment=[s]{/*}{*/},
  columns=fullflexible,
  keepspaces=true,
  sensitive=true
}

\newcounter{tamarinrule}
\newcounter{tamarinlemma}
\renewcommand{\thetamarinrule}{\arabic{tamarinrule}}
\renewcommand{\thetamarinlemma}{\arabic{tamarinlemma}}

\newcommand{\ruleheader}[1]{
  \refstepcounter{tamarinrule}
  \label{rule:\thetamarinrule}
  \textbf{\underline{Rule~\thetamarinrule:} #1}\\
}

\newcommand{\lemmaheader}[1]{
  \refstepcounter{tamarinlemma}
  \label{lemma:\thetamarinlemma}
  \textbf{\underline{Lemma~\thetamarinlemma:} #1}\\
}

\lstdefinestyle{tamarinstyle}{
  language=Tamarin,
  basicstyle=\footnotesize\ttfamily,
  frame=single,
  breaklines=true,
  commentstyle=\color{gray},
  escapechar=|,
}

\begin{document}

\title{\acron: Lightweight Enclave Isolation with Prismatic Capabilities }

\ifnotanonymous
\author{Merve G\"{u}lmez\textsuperscript{\equalContribution,\affiliationA}\thanks{\textsuperscript{\equalContribution}Both authors contributed equally to this research.},
Adam Caulfield\textsuperscript{\equalContribution,\affiliationB},
Håkan Englund\textsuperscript{\affiliationA},
N. Asokan\textsuperscript{\affiliationB,\affiliationC}, Thomas Nyman\textsuperscript{\affiliationD}\\
\textit{\textsuperscript{\affiliationA}Ericsson Security Research, \textsuperscript{\affiliationB}University of Waterloo,}\\\textit{ \textsuperscript{\affiliationC}KTH Royal Institute of Technology, \textsuperscript{\affiliationD}Ericsson Product Security} \\
\{merve.gulmez,hakan.englund,thomas.nyman\}@ericsson.com, \\acaulfield@uwaterloo.ca, asokan@acm.org}
\fi

\ifnotarxiv
\IEEEoverridecommandlockouts
\makeatletter\def\@IEEEpubidpullup{6.5\baselineskip}\makeatother
\IEEEpubid{\parbox{\columnwidth}{
		Network and Distributed System Security (NDSS) Symposium 2027\\
		22--26 March 2027, Seoul, Republic of Korea\\
		ISBN 978-1-970672-09-1\\  
		https://dx.doi.org/10.14722/ndss.2027.[23$|$24]xxxx\\
		www.ndss-symposium.org
}
\hspace{\columnsep}\makebox[\columnwidth]{}}
\fi

\maketitle
\begin{abstract}
\Glspl{tee} protect sensitive code and data from external interference, but lack inherent memory safety. CHERI can enforce spatial memory safety at the object level. Attempts to establish a TEE using CHERI primitives suffer from (1) expensive capability revocation operations, (2) need to rely on the host \gls{os} to support provenance tracking and physical memory protection, (3) expensive domain transitions, and (4) lack of support for remote attestation.

We introduce \emph{\enclavecaps} and present \acron, a TEE architecture for CHERI leveraging prismatic capabilities to create userspace enclaves while addressing these challenges. \acron binds enclave `\capcolors' (identifiers) in \enclavecaps to physical memory access controls, enables O(1) ownership establishment without memory sweeps, and supports efficient domain transitions that atomically activate and deactivate \enclavecaps. Additionally, \acron enables remote attestation of its enclaves. We demonstrate that execution of userspace enclaves in \acron incurs only moderate overhead ($\leq15\%$), a significant improvement over the same workloads under Intel SGX. 
\end{abstract}

\ifnotarxiv
\IEEEpeerreviewmaketitle
\fi
\glsresetall

\section{Introduction}\label{sec:intro}

Hardware-isolated \emph{\glspl{tee}} have become a cornerstone of confidential computing, enabling applications to protect sensitive code and data even when the host \gls{os} is compromised or malicious. \glspl{tee} provide an isolated computing environment within which code and data are inaccessible to the host system, except when authorized and through well-defined interfaces. The use of \glspl{tee} in industry~\cite{trustzone,amd-sev-snp,sgx,tdx} and academia~\cite{sancus,keystone,sanctum} is ubiquitous across a broad range of systems.
\Glspl{tee} protect their code and data from direct manipulation. However, the Achilles' heel of \glspl{tee} is the threat surface exposed by the software within; security lapses in real-world \glspl{tee} are predominantly caused by vulnerabilities in the trusted software, such as memory-safety issues~\cite{Shen15,Laginimaineb16,Laginimaineb16a}. The memory safety of TEE code is typically not enforced by the architecture itself. 
It can be achieved though memory-safe languages, e.g., Rust, but \gls{tee} \glspl{os} or their trusted runtime typically rely on low-level C (or C++) components repurposed from mainstream toolchains.
The threat surface is exacerbated by bloated \glspl{tcb}; the software threat surface of a \gls{cvm} is hardly any smaller than that of its non-confidential counterpart, unless explicitly minimized as part of system hardening.

\gls{cheri}~\cite{Watson23a} guarantees spatial memory safety, extending conventional \glspl{isa} with hardware \emph{capabilities} that embed object types, bounds, permissions, and validity tags into pointers themselves.
\Gls{cheri} naturally enforces spatial memory safety at the object level and enables fine-grained compartmentalization through capability sealing and controlled invocation~(\Cref{sec:cheri}). 

While \gls{cheri}'s per-pointer access control and sealed invocation are relevant to userspace \glspl{tee}, also known as \emph{enclaves}, \gls{cheri} was not built to enforce \emph{exclusive access} of capabilities belonging to a \gls{tee}.
CHERI-TrEE proposes a \gls{cheri}-based TEE design ~\cite{van2023cheri}; however, it requires a linear memory sweep to prove exclusive ownership, and further operates directly on physical memory without virtual-address support---precluding deployment on systems with an OS-managed MMU.
Capstone~\cite{Yu23} avoids sweeps by introducing linear capabilities with revocation trees, but imposes tree-traversal overhead and requires a bespoke compiler toolchain.
\gls{cheri}, by design, avoids such indirection: once a capability is derived, no historical lookup table tracks where it propagates.
This makes ensuring exclusive access in \gls{cheri} fundamentally difficult---there is no efficient way to revoke all capabilities overlapping a given region short of scanning all memory and registers.
Virtual-to-physical address translation further exacerbates this problem: capabilities operate on virtual addresses, yet enclave isolation must protect physical memory.

\textbf{This paper and contributions.}
We propose \emph{prismatic capabilities} for the \gls{cheri} architecture to enable lightweight enclave execution in userspace, secure from even an untrusted operating system.
A \enclavecap is a new capability type for enclaves which receive a \emph{\capcolor} (provenance identifier for an enclave instance), and where data additionally receives a \emph{\gradient}\footnotemark (provenance identifier for dynamic allocations). \Capcolors are bound to an assigned physical memory region and simultaneously gate memory accesses to provide exclusive access, control invocation, and serve as an attestation identity.
\Capcolor assignment and life-cycle are managed exclusively by a machine-mode (\acrshort{m-mode}) \gls{tcb}, ensuring that no software below machine mode---including the \gls{os} kernel---can assign and retire enclave \capcolors.
Hardware call gates enable safe, zero-trap context switches into and out of a running enclave.
Finally, a hardware-\capcolor{}-addressed table enforces access control at the physical layer, ensuring that even a malicious OS that controls virtual address translation cannot reach enclave memory.
\footnotetext{We distinguish between \capcolors and \gradients as, as explained in~\Cref{sec:highlevel}, \enclavecaps combine both to provide isolation and temporal safety.}
In summary, our contributions are:
\begin{itemize}[nosep, leftmargin=*]
    \item The design of prismatic capabilities (\Cref{sec:highlevel}): a \gls{cheri} extension with capability-based physical memory isolation, hardware call gates for trap-free enclave transitions, and minimal M-mode \gls{tcb}  (\glsdesc{hmm}) for life cycle management and remote attestation.
    \item \acron: a realization of prismatic capabilities on the CHERI-RISC-V QEMU full-system emulator and in an \gls{fpga} softcore based on the CHERI-Toooba~\cite{CTSRD24a} processor (\Cref{sec:cam-hw}).
    \item Software support for \acron through a CheriBSD kernel driver (\Cref{sec:kernel}) and userspace enclave \gls{sdk} that supports source-compatibility\footnotemark{} with \gls{sgx} enclaves (\Cref{sec:sgx-apps}).
    \item Demonstrating the efficacy of \acron through three case studies using real-world \gls{sgx} enclaves (\Cref{sec:usecases}) recompiled to \acron with moderate run-time overhead ($\approx6--15\%$).
    \item Proof of security for \acron's attestation protocol through formal modeling and symbolic protocol verification (\Cref{sec:security}).
\end{itemize}
\smallskip
We will make the full source code and \gls{rtl}-implementation of \acron available for artifact evaluation and plan to open-source all the artifacts on acceptance. 

\footnotetext{With the exception of source-code changes necessary to make the enclave-code conformant to \gls{cheri} C/C++ and the recompilation of enclave code (typically targeting 64-bit x86) to our \acron-enabled CHERI-Toooba target.}

\section{Background}\label{sec:background}

\subsection{CHERI}\label{sec:cheri}

\begin{figure}[t]
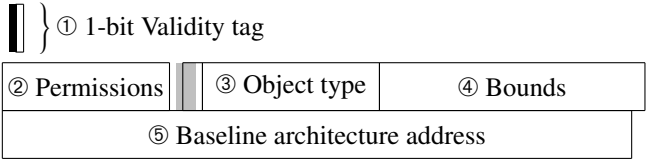

    \begin{bytefield}[bitwidth=0.3em]{1}
        \begin{rightwordgroup}{\dCOne{} 1-bit Validity tag}
            \colorbitbox{black}{1}{}
        \end{rightwordgroup} \\
        \bitbox[]{1}{} 
        \vspace{-1.5em}
    \end{bytefield}\\
    \begin{bytefield}[bitwidth=0.37em]{64}
        \bitbox{17}{\dCTwo{} Permissions} & \colorbitbox{lightgray}{2}{} &
        \bitbox{18}{\dCThree{} Object type} & \bitbox{27}{\dCFour{} Bounds} \\
        \bitbox{64}{\dCFive{} Baseline architecture address} \\
    \end{bytefield}
    \vspace{-1.5em}
    \caption{\gls{cheri} capability layout from Watson et al.~\cite{Watson19}}\label{fig:chericap}
\end{figure}

\gls{cheri} is an \gls{isa} extension that adds a capability-based hardware-software co-design for memory protection, using hardware-supported capabilities to enforce \emph{spatial safety} for pointers.
The \gls{cheri} \gls{isa} specification~\cite{Watson23a} defines how capabilities are represented in registers and memory and the instructions that manipulate them safely. 

As shown in \Cref{fig:chericap}, a \gls{cheri} capability is twice the width of a native pointer type---128~bits on 64-bit platforms---plus a separate validity-tag bit \dCOne held outside the addressable data to prevent tampering by non-capability-aware instructions.
Capability-aware instructions preserve this tag, but it is cleared by unauthorized manipulation or attempts to inject arbitrary capabilities.
\gls{cheri} widens the general-purpose and address registers to hold full capabilities, each comprising:
\begin{itemize}[leftmargin=*, nosep]
    \item \textbf{Permissions} \dCTwo: a bitmask defining allowed operations.
    \item \textbf{\Gls{otype}} \dCThree: a signed integer enabling temporary ``sealing'' and ``unsealing'' of capabilities for opaque pointers and fine-grained in-process isolation.
    \item \textbf{Bounds} \dCFour: The valid memory range relative to the baseline architecture address \dCFive, using a compressed encoding~\cite{Woodruff19} to reduce the space taken by a capability at the cost of stricter alignment for larger object allocations.
\end{itemize}
Special-purpose registers, e.g. the \gls{pc} and \gls{sp}, are replaced with equivalent capability-registers: \gls{pcc} and \gls{csp}.
\gls{cheri} enforces spatial safety by associating each allocation with a capability describing its valid range and permissions.
New capabilities are derived from existing ones, while maintaining \emph{monotonicity} that ensures that new capabilities cannot exceed the permissions or bounds of their parent. 
The lineage is traceable to initial boot-time capabilities; the \gls{dcc} and \gls{pcc}.
\gls{cheri}'s sealing enables compartmentalization, rendering sealed capabilities inert until passed to a \texttt{cinvoke} instruction which jumps to an entry point within a compartment and simultaneously unseals a data capability, granting access to compartment data.
Exception handling uses the same mechanism for limited non-monotonicity.
Extensions to \gls{cheri} have also explored sandboxing~\cite{Chisnall17}, initialization safety~\cite{Georges21,Gulmez25a, wang2026}, safe speculation~\cite{Fuchs23, Fuchs24}, and side-channel resistance~\cite{ElAtali25}.

\paragraph{Capability relocation.}
Capabilities cannot be embedded statically in an executable: storing a pointer-sized value as ordinary data lacks a valid tag, and conventional relocation types cannot describe a capability's base, bounds, and permissions.
The \gls{cheri} toolchain emits a \gls{caprelocs} with one entry per capability to be initialized at program launch, recording the slot to fix up together with the target object's base, offset, and size.
The run-time linker---or, for statically linked binaries, equivalent startup code embedded in the C runtime---walks this table and, for each entry, derives a corresponding capability from the appropriate root, sets its tag, and writes it into the corresponding pointer slot, so that globals, function pointers, and similar values hold valid capabilities before application code runs.
Because every capability originates this way, the monotonicity guarantee extends from these roots to all capabilities in the process~\cite{Davis19}.

\paragraph{\gls{cheri} temporal-safety enforcement.}
By design, CHERI avoids indirection such as  table-based lookups. Temporal safety therefore relies on a software mechanism that periodically sweeps memory and revokes capabilities~\cite{WesleyFilardo20,Filardo24}. \PICASSO~\cite{Gulmez26} introduces colored capabilities to track allocation provenance and shows that single-bit indirection accelerates temporal safety in CHERI.
\CCs decouple the validity of a capability's provenance (the allocation it represents) from the validity of the capability itself.
Each colored capability carries a \ccpid encoded in the \texttt{otype} field (referred to as the capability's \emph{"color"}).
A corresponding \gls{pvb} is stored separately in a hardware-managed \gls{pvt} in process memory, indexed by the \ccpid.
Hardware checks the \gls{pvb} on every load or store issued through a colored capability.
Dereferencing \ccs without a valid \gls{pvb} bit is prevented.
This allows all capabilities sharing provenance to be retracted at once by invalidating their \gls{pvb}, disabling dereferencing dangling \ccs without permanently revoking them.

\subsection{Trusted Execution Environments}

\Glspl{tee} are a class of hardware extensions that provide a protected runtime isolated from a system's host environment: code and data in a \gls{tee} can only be directly called, read and written to by code in the same \gls{tee}. This is enforced on an architectural level, so even privileged software on the system (e.g., host operating system) cannot directly access code/data belonging to a \gls{tee}. The nature of the trusted environment varies based on the particular \gls{tee} implementation. 

Many designs of \glspl{tee} are available in commercial \glspl{cpu}~\cite{sgx,trustzone,tdx,amd-sev-snp,arm-realms}. For example, Intel's SGX~\cite{sgx} enables \glspl{tee} in userspace (referred to as enclaves) in Intel processors.  ARM TrustZone~\cite{trustzone} partitions all system resources (e.g., timers, peripherals, interrupts) between the host and \gls{tee} (referred to as the \textit{normal} and \textit{secure world}, respectively) in ARM Cortex-M and -A processors. Additionally, ARM TrustZone enables a set of \textit{trusted applications} to execute within the bounds of its \gls{tee} (atop a small \textit{trusted operating system} in the Cortex-A variant). Recent commercial CPUs have also enabled virtual machines inside a \gls{tee} ~\cite{tdx,amd-sev-snp,arm-realms}, protected from the host virtual machine manager.

Academic proposals for \gls{tee} design have been prevalent in recent years~\cite{sanctum,timber-v,cure,penglai,dorami,sancus,flicker,sanctuary,trustlite} including those targeting the RISC-V \gls{isa}~\cite{sanctum,timber-v,cure,penglai,dorami}. Among them is Keystone~\cite{keystone}, which enables construction of SGX-like \glspl{tee} using RISC-V primitives, leveraging the \gls{pmp} mechanism to restrict access to \gls{tee} memory. It also establishes a secure monitor running in machine mode to act as a reference monitor that controls the \gls{pmp} configuration at run time.
Regardless of their specifics, \glspl{tee} guarantee hardware-backed isolation from the host \gls{os}. Because of this, \glspl{tee} can enable remote attestation, a challenge-response protocol in which a prover (\prv) aims to convince a verifier (\vrf) that their software/hardware configuration is valid. Remote attestation requires the prover to have a \textit{root of trust} on their device that can measure and report on the software/hardware components of interest to \vrf. The latter is typically conducted by computing a message authentication code (MAC) or digital signature over the components of interest, thus requiring secure key storage and usage. Due to this requirement, the hardware-backed security properties of \glspl{tee} naturally constitutes such a root of trust, and some even build attestation operations directly into the \gls{cpu} fabric~\cite{sgx,amd-sev-snp,tdx}.

\section{Problem Statement}

\subsection{System and Adversary Model}\label{sec:threatmodel}
We consider a \prv system running a host \gls{os} to manage the execution of multiple userspace processes, provide inter-process isolation, and perform virtual-to-physical translation. We assume a \gls{cpu} core with a memory management unit. We also consider a CPU equipped for CHERI, such that the entire runtime environment can operate in \emph{\glshyph{purecap} mode} (i.e., all pointers are turned to capabilities and non-capability memory accesses are prohibited).
Prismatic capabilities aim to support establishment of enclaves, housing code and data that is isolated from the most privileged software on the system. 
We consider an \gls{adv} in line with related works~\cite{keystone,sanctum}, which we assume can exploit vulnerabilities in the \gls{os} to gain control and call an arbitrary sequence of operations at its disposal. With this control, \gls{adv} can modify memory not explicitly protected to either (1) leak secrets (2) tamper with an enclave's code/data. 

We assume \gls{cheri}'s spatial-safety properties hold as intended. As such, we assume permissions and bounds can only be \emph{decreased} through the creation of a new (equal- or less-privileged) capability (cf. \gls{cheri} monotonicity in \Cref{sec:cheri}).

An enclave may divulge \enclavecaps pointing to enclave memory to untrusted software.
In \Cref{sec:highlevel} we explain how \gls{cpu} hardware ensures such capabilities remain opaque to untrusted software and other enclaves.
However, we assume enclave software to be responsible for validating that any \enclavecaps it receives from untrusted software are pointing to valid and correct objects within the enclave.
This assumption is consistent with \glspl{tee} which must validate any untrusted input from outside their trust boundary before the input is used in sensitive operations.
Hardware attacks that require physical access to circumvent hardware protections (e.g., hardware-protected code/data) are out-of-scope for this work, as they require a set of mitigations that are orthogonal to the focus of this work~\cite{louka2026pmplease, henes2026, demeulemeester2026batteringram, demeulemeester25badram}. Consistent with prior TEE work~\cite{sgx, Yu23, van2023cheri, sanctum}, availability is out of scope: since \gls{adv} controls the \gls{os}, it can trivially deny service by refusing to schedule the host process. Regarding remote attestation, we also assume \adv may discard, inject, or modify network messages~\cite{Dolev83}.
We exclude microarchitectural side-channels from our adversary model, but discuss this threat surface in-detail in Appendix~\ref{apdx:discussion}.

\subsection{Design Goals and Challenges}\label{sec:goals-and-challenges}

Prior work~\cite{Yu23,van2023cheri,Gulmez26} has extended traditional usage of CHERI capabilities beyond spatial safety. However, the following challenges must be addressed to enable an efficient capability-based \gls{tee} design:
\begin{enumerate}[nosep,leftmargin=*]
    \item obtaining exclusive ownership of capabilities belonging to the \gls{tee} without performing expensive look-up operations (e.g., memory sweeps or tree-traversal);
    \item enabling privilege separation for management of metadata used to track provenance of TEE; 
    \item enabling exclusive access with an untrusted host \gls{os}, as capabilities are specified on virtual addresses and enclave protection needs to apply to physical addresses;
    \item enabling efficient domain transitions, such that enter and exit from enclave does not require performing an expensive context-switch routine to disable an enclave's capabilities;
    \item enabling external verifiers to assess whether  a particular enclave was created using the expected code/data and executes on the anticipated system. 
\end{enumerate}
To address these challenges, a capability-based \gls{tee} architecture must have the following properties: 
\begin{enumerate}[label=\textbf{[P\arabic*]}, nosep]
    \item \textbf{Exclusive Ownership Management}\label{prop:exclusiveownership}: 
    assigning and revoking exclusive ownership of an enclave's capabilities must not require expensive search operations, and metadata for provenance tracking must be inaccessible to untrusted software. \ref{prop:exclusiveownership} ensures enclave isolation from its untrusted host userspace process.
    \item \textbf{Physical Memory Exclusivity}\label{prop:exclusivememory}: enclave capabilities must be bound to physical access control measures, in addition to out-of-the-box spatial safety on virtual addresses. \ref{prop:exclusivememory} ensures enclave isolation from the untrusted host \gls{os}.
    \item \textbf{Safe Domain Transitions}\label{prop:domaintransitions}: entering and exiting an enclave (calls/returns or interrupts by the untrusted \gls{os}) must not enable tampering with enclave state.
    \item \textbf{Spatial and Temporal Safety}\label{prop:full-mem-safety}: full spatial and temporal safety across both enclave and non-enclave execution.
    \item \textbf{Secure Measurement and Attestation}\label{prop:attestation}: enclave code and data is cryptographically measured upon creation, with measurements stored in \gls{tcb} memory; attestation reports bind the enclave to both the installed \gls{tcb} and device. 
\end{enumerate}

\begin{figure*}[ht!]
    \centering
    \includegraphics[width=0.75\textwidth]{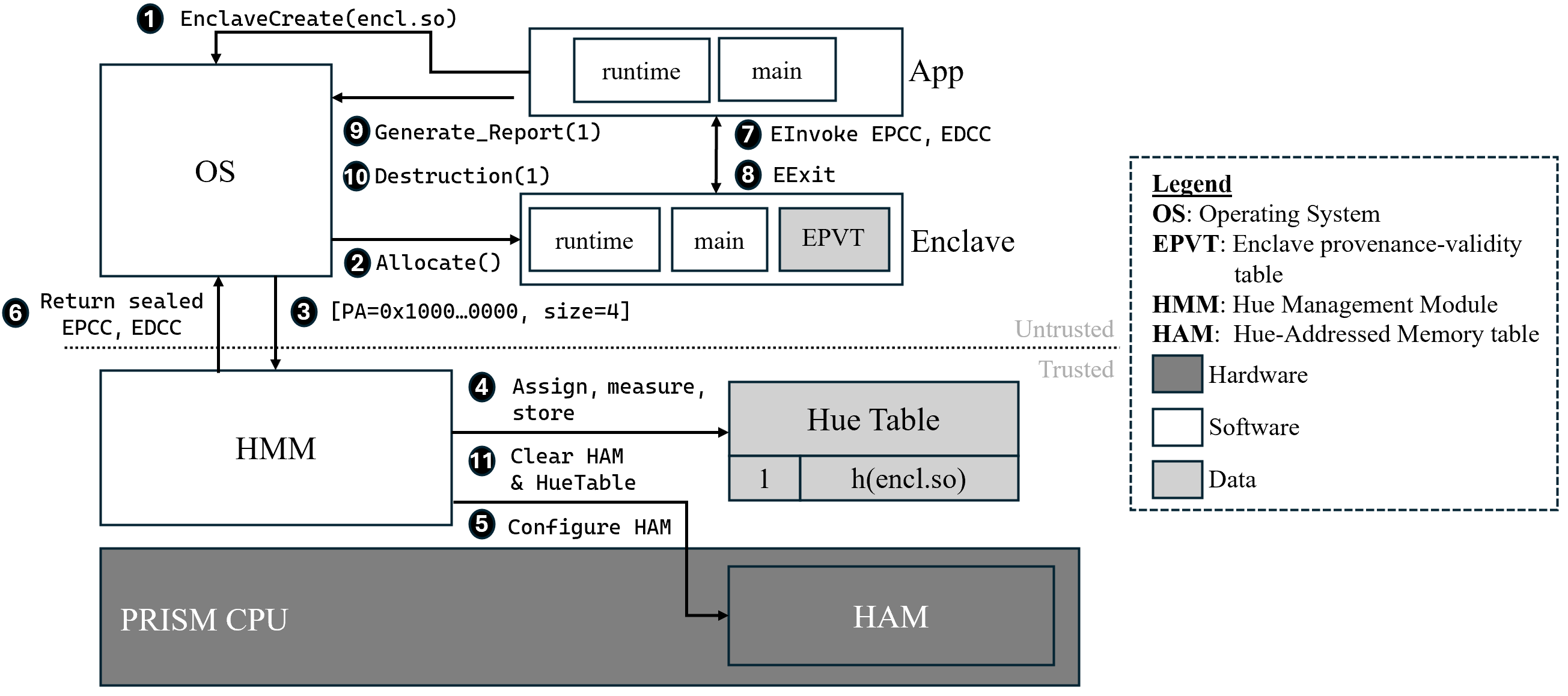}
    \caption{Overview of \emph{\enclavecaps}: hardware/software components and the enclave life cycle.}
    \label{fig:life-cycle}
\end{figure*}

\section{Prismatic Capabilities Design}\label{sec:highlevel}
Prismatic capabilities extend \gls{cheri} capabilities with a \emph{\capcolor-based}\footnotemark isolation mechanism that enforces exclusive access to physical memory regions. The name draws on the analogy of an optical prism separating white light into distinct spectral components: a single address space is partitioned into isolated protection domains, each identified by a distinct \capcolor and accessible only      
through capabilities bearing the same \capcolor.
Prismatic data capabilities additionally carry a \gradient for temporal safety protection of dynamically allocated memory regions within a hue (enclave instance).
\Cref{fig:life-cycle} shows the components of a prismatic capability system and depicts the enclave life cycle. 

\footnotetext{A ``\emph{\capcolor}'' contrasts with the use of ``\emph{color}'' used to refer to subject or object identifier in tag-based memory access-control schemes or by \PICASSO with respect to \gls{cheri}'s \gls{otype} while encoding a \ccpid (\Cref{sec:cheri}).}

A \enclavecap carries a \emph{\capcolor attribute} (encoded in the capability's object type field) assigned by a machine-mode \gls{tcb} at enclave creation time.
Each \enclavecap belonging to a particular enclave instance carries the same \capcolor.
At each memory access, \gls{cpu} hardware compares the capability's \capcolor against entries in a hardware-managed \emph{\capvalidtable} table. 
The \capvalidtable associates contiguous physical memory regions with the enclave's \capcolor identifiers.
The check applies to both instruction fetch and data
load/store (keyed on the authorizing capability); additionally, for data loads/stores, a check on the validity of the \gradient's provenance is performed. The memory access succeeds if
\begin{inparaenum}[1)]
\item the access's physical address, after translation, resides within the enclave's physical region,
\item the \capcolor and \gradient of the capability authorizing the access matches that registered for the target region, and 
\item the access is otherwise valid according to the bounds, permissions, and other invariants enforced by \gls{cheri}.
\end{inparaenum}
Crucially, \enclavecaps are usable only from within the enclave with the same \capcolor \ref{prop:exclusiveownership}.
Additionally, even if stale or forged (non-\capcolor) capabilities reference the same physical addresses, they cannot access a \capcolor-locked region \ref{prop:exclusivememory}.
Together, \ref{prop:exclusiveownership} and \ref{prop:exclusivememory} provide isolation without page-table isolation or software-mediated domain crossings.

\gls{cheri}'s monotonicity property ensures any capability derived from a prismatic capability shares the original capability's \capcolor, but capabilities of an arbitrary \capcolor cannot be forged.

\subsection{System Components}
As shown in \Cref{fig:life-cycle}, the system is comprised of hardware components, software modules, and data structures. Among the software modules are the following:
\begin{itemize}[nosep,leftmargin=*]
    \item \textbf{\gls{os}:} the system's full feature-rich host \gls{os}. The \gls{os} is assumed to be compromised and acting maliciously.
    \item \textbf{App}: a host application, which requests the creation of an enclave within its own address space.
    \item \textbf{Enclave}: a subsection of the App's address space used for operations isolated from the rest of the system. 
    \item \textbf{\tcb}: a security monitor that operates at machine mode level, and is responsible for crucial aspects of an enclave's life cycle.
\end{itemize}

We assume there are just two main hardware components: 
\begin{itemize}[nosep,leftmargin=*]
    \item \textbf{\acron CPU}: the \gls{cpu} core which is capable of enforcing standard CHERI spatial memory safety, augmented with support for \enclavecaps.
    \item \textbf{\gls{cam}}: \acron's hardware mechanism for enforcing physical memory isolation for enclaves (full details in Section~\ref{sec:cam-hw}). It is a sub-component of the \acron CPU.
\end{itemize}
Finally, the \tcb maintains one critical data structure called the \emph{\caphashtable}, which maintains a mapping of enclave \capcolors and the measurement of the code and data upon creation. 

\subsection{Life-cycle}
We describe the enclave life cycle stages: creation, execution (invocation, return, preemption), attestation, and destruction:

\textbf{Enclave Creation.} In \circled{1}{black}{white}, the host application creates an enclave by calling the \gls{os} and passing the enclave binary (e.g., \texttt{encl.so} in \Cref{fig:life-cycle}). In \circled{2}{black}{white}, the \gls{os} allocates a contiguous physical region and invokes the \tcb (via a privileged call) in \circled{3}{black}{white}. In \circled{4}{black}{white} the \tcb
assigns a unique \capcolor to the enclave instance (e.g., 1 in the example depicted in \Cref{fig:life-cycle}), measures the memory contents by computing a cryptographic hash over the enclave's code and data, and records the \capcolor-to-measurement binding in \caphashtable in \tcb-private storage.
\Cref{fig:life-cycle} shows a mapping between the \capcolor 1 and \texttt{h(encl.so)} in the \caphashtable.

In \circled{5}{black}{white}, \tcb updates the \gls{cam} to associate the \capcolor with the physical region (further details in Section~\ref{sec:cam-hw}). Then, it locks the memory against all memory accesses except for accesses through \enclavecaps with the associated \capcolor.
Finally, the \tcb seals a code capability and a data capability (e.g., \texttt{EPCC} and \texttt{EDCC} in \Cref{fig:life-cycle}) with the
assigned \capcolor, and returns them to the host application in \circled{6}{black}{white}. After creation, the sealed capabilities are the only means to
enter the enclave. \Capcolor accesses are confined: a capability carrying a TEE \capcolor may only access memory explicitly assigned to that \capcolor; the host \gls{os} cannot redirect the enclave to attacker-controlled memory by manipulating page tables.

\textbf{Enclave Execution.}
The enclave is invoked in \circled{7}{black}{white} by the host application. It enters the enclave via a single unprivileged hardware instruction (described further in Section~\ref{sec:impl}). Upon the execution of this instruction, the \gls{cam} validates that both code and data capabilities carry matching \capcolor, activates the corresponding \capcolor entry,  installs the code capability, and transfers control to the enclave entry point. A return sentry is
saved for later use by the enclave to return to the caller.

As a result of this design, no trap to the \tcb occurs; invocation is entirely a hardware operation. From this point, instruction fetches and data accesses succeed because the code and data capabilities carry matching \capcolor. Any code whose capabilities carry a
different \capcolor is denied by the \gls{cam}. 

Interrupts may fire during enclave execution. The \tcb 
preempts the enclave by saving its register state into
enclave memory (inaccessible to the \gls{os}), deactivates the \gls{cam} entry to
re-lock the enclave's memory, scrubs the trap frame, and redirects execution to
\gls{os} interrupt handler. The \gls{os} processes the interrupt, unaware that an enclave
was executing. To resume, the host simply re-invokes the enclave; entry
logic detects the saved state and the \tcb restores execution from the interrupted point.  Synchronous faults (such as \gls{cheri} exceptions) are handled similarly, except they terminate the enclave execution rather than saving resumable state. 

In \circled{8}{black}{white}, the enclave exits via a single unprivileged instruction that atomically deactivates the \capcolor (re-locking the enclave's memory) and returns control
to the caller. The \gls{cam} entry remains valid, so the enclave can be re-entered at a later time.

\textbf{Attestation}. The host application responds to remote attestation requests by calling the OS to transfer control to the \tcb for report generation in \circled{9}{black}{white}. The \tcb constructs a report, containing measurements of the enclave code, the \capcolor, an attestation challenge, and measurement of the \tcb code. This report is then signed using the device signing key, binding the measurements of the enclave and \tcb at the time of the attestation request to the particular device. The remote attestation protocol is discussed in more details in \Cref{sec:ra}.

\textbf{Destruction.}
In \circled{10}{black}{white}, the host application can permanently destroy an enclave, by making such a request to the  
\gls{os}, which in return forwards the request to the \tcb via a privileged call.
The \tcb invalidates the \capcolor in the \gls{cam}, \emph{retires} the \capcolor, then clears and releases the physical memory in \circled{11}{black}{white}. \Enclavecaps become unusable when the \capcolor is retired: their \capcolor no longer corresponds to a \gls{cam} entry, so subsequent invocation attempts fault.
Retired \capcolors can later be reclaimed by the \tcb by performing a system-wide sweep to revoke stale \enclavecaps held by untrusted software.
This \capcolor-revocation is independent of the enclave life cycle.

\subsection{Enclave temporal safety}

While enclave spatial safety follows from the baseline \gls{cheri} design (\Cref{sec:threatmodel}), we adapt \cchyph-based temporal-safety enforcement (\ref{prop:full-mem-safety}) from \PICASSO~\cite{Gulmez26} (\Cref{sec:cheri}).
Our adversary model precludes storing provenance-validity data in a host-controlled \emph{\gls{pvt}}.
Instead, each enclave contains an \emph{\gls{epvt}}.
\Enclavecaps reserve a portion of \gls{cheri}'s \gls{otype} as \capcolors.
The remaining \glspl{otype} are treated as \PICASSO \gradients.

The \gls{epvt} contains 1-bit \emph{\gls{pvb}} entries indexed by the \gradient (i.e., \ccpid).
On allocation, a trusted memory allocator within the enclave assigns a unique \gradient to \enclavecaps and marks
the corresponding \gls{pvb} in the \gls{epvt} \emph{\ccactive}.
On free, the allocator marks the corresponding \gls{pvb} \emph{\ccinactive}.
All loads or stores with a \gradient are subject to an additional hardware check against the \gls{epvt} in parallel to \capcolor checks.
Once a \gradient has been invalidated, the \gls{cpu} ensures that all capabilities with that \gradient cannot be dereferenced, preventing \gls{uaf} and \gls{uar} conditions within the enclave.
The revocation sweep which invalidates stale heap capabilities and reclaims \gradients (performed by a separate kernel thread in \PICASSO) is performed by the trusted runtime inside the enclave. 
Unlike \gls{cheri} revocation sweeps, enforcing temporal safety within enclave instances only requires sweeping the particular enclave memory.
According to our system and adversary model (\Cref{sec:threatmodel}), enclave code is responsible for validating any capability divulged to, and later received back from untrusted code.
Since temporal-safety enforcement does not extend to capabilities that have left the enclave boundary (due to our adversary model~\Cref{sec:threatmodel}), enclave software must ensure potentially stale capabilities originating from outside the enclave are not used.

\section{Implementation Details}\label{sec:impl}
In this section, we present \acron, our implementation of prismatic capabilities for the CHERI-RISC-V architecture. We implemented two versions of \acron: 1) a QEMU-system-
CHERI128 full-system emulator, and 2) a softcore based on the
CHERI-Toooba processor \gls{ip} (\Cref{sec:cam-hw}). We also implement an M-mode security monitor (\gls{hmm}) (\Cref{sec:tcb}), a CheriBSD kernel
driver (\Cref{sec:kernel}), and an SGX-compatibility layer comprising a \gls{trts} and an \gls{urts}\footnotemark (\Cref{sec:sgx-apps}).

\footnotetext{The \gls{trts} and \gls{urts} mirror their counterparts in the \gls{sgx} runtime.}

\subsection{\acron CHERI-RISC-V Implementation}\label{sec:cam-hw}

\paragraph{Hue-Addressed Memory.} The \capvalidtable is the hardware mechanism that enforces enclave physical memory
isolation. It is a 64-entry fully-associative table positioned between the \gls{mmu} and
the memory subsystem. On every memory access (instruction fetch or data load/store), after the \gls{mmu} produces a physical
address, the \capvalidtable checks whether the capability used for the access falls within a protected region
and, if so, whether it carries the correct
\capcolor. Where the \gls{mmu} enforces virtual-to-physical translation
and page-level permissions, the \capvalidtable enforces domain separation: it determines
whether the code currently executing has the right to access a particular
physical address range, based on the \capcolor carried by the capability being
used for the access. 
\capvalidtable entries use \gls{napot} encoding, storing a base
page number and a size exponent rather than explicit base and top addresses.
This replaces two magnitude comparators with a single masked equality check. Each entry totals 66~bits:

\begin{center}
\footnotesize
\begin{tabular}{lrl}
\toprule
\textbf{Field} & \textbf{Bits} & \textbf{Description} \\
\midrule
\texttt{basePage}  & 52 & physical page number of region start \\
\texttt{log2Pages} &  6 & region size $= 2^{\texttt{log2Pages}}$ pages \\
\texttt{hue}     & 6 & enclave instance identifier \\
\texttt{valid}     &  1 & entry is programmed \\
\texttt{active}    &  1 & enclave is currently executing \\
\bottomrule
\end{tabular}
\end{center}

For a given access to page number $P$ and entry with base $B$ and exponent $L$, the entry matches when $(P \mathbin{\&} \mathord{\sim}(2^L - 1)) = B$.
All 64~entries are evaluated in parallel.
If one or more entries match, access is granted only if the matching entry is
active \emph{and} the capability's \capcolor equals the entry's \capcolor. The same check gates both instruction fetch and data access, each on its
post-translation physical address: a fetch is checked against the \capcolor of the
executing \gls{pcc}, and a load or store against the \capcolor of the capability
authorizing it. Because the check happens after address translation, a
malicious kernel cannot use the page tables to redirect an enclave's fetches or
data accesses to physical memory of another \capcolor (\ref{prop:exclusivememory}, \Cref{sec:goals-and-challenges}). The mismatched \capcolor is
denied regardless of how the virtual address was mapped.
M-mode accesses are unconditionally exempt, allowing the \gls{hmm} to manage enclave
memory regardless of \gls{cam} state. Each entry additionally holds the temporal-safety metadata needed to verify the \gradient, as described in \Cref{sec:temporal-safety}, the enclave-virtual base address of its \gls{epvt} bitmap and the number of validity slots it covers. 
The \capvalidtable is core-local (similar to RISC-V's \gls{pmp}).
In a multi-core configuration the \tcb must program the corresponding entry on every core. In this work, we limit \acron to a single core.

\paragraph{Domain Transition.}
We introduce \EInvoke, an extension to the \texttt{cinvoke} instruction, which activates the corresponding \gls{cam} entry (i.e., sets its \texttt{active} bit) when the sealed \gls{otype} falls within the \gls{tee} \capcolor range, before transferring control to enclave code. The \EInvoke instruction remains unprivileged and does not trap to the \tcb.
We also introduce a new unprivileged instruction (\EExit) used to exit from an enclave.
This instruction reads the \capcolor from the current \gls{pcc}, deactivates the corresponding \gls{cam} entry (i.e., clears the \texttt{active} bit), and jumps to the return sentry. The enclave's memory is immediately re-locked upon return via \EExit. Before the exit, the \gls{trts} scrubs general-purpose register, so no enclave secret or capability remains in registers when control returns to untrusted code; the same scrub guards the \gls{ocall} exit path. The \gls{trts} and \gls{urts} are discussed further in \Cref{sec:sgx-apps}.

On an out-of-order core, enclave entry must be serialized: a \gls{cam} entry activates only when \EInvoke commits, so \acron suppresses its speculative redirect and, at commit, activates the entry and squashes younger instructions before fetching enclave code. As a result, no instruction runs under the enclave's \capcolor before the \capcolor is active (\ref{prop:domaintransitions}, \Cref{sec:goals-and-challenges}). \EExit needs no barrier, as it returns to host code without a \capcolor.

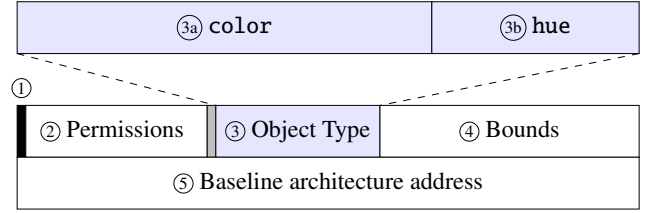
\begin{figure}[t]
    \resizebox{\columnwidth}{!}{
    \begin{tikzpicture}
        \filldraw[fill=blue!10, draw=black] (0, 2.5) rectangle (6,3.25);
        \node[] at (3, 2.875) {\circled{3a}{blue!10}{black} \texttt{color}};

        \filldraw[fill=blue!10, draw=black] (6, 2.5) rectangle (9,3.25);
        \node[] at (7.5, 2.875) {\circled{3b}{blue!10}{black} \texttt{hue}};
        \draw[dashed] (0, 2.5) -- (2.875, 1.75);
        \draw[dashed] (9, 2.5) -- (5.25, 1.75);
        \node[] at (0.0625, 2) {\circled{1}{white}{black}};
        \filldraw[fill=black, draw=black] (0, 1) rectangle (0.125,1.75);
    
        \filldraw[fill=white, draw=black] (0.125, 1) rectangle (2.75,1.75);
        \node[] at (1.375, 1.375) {\circled{2}{white}{black} Permissions};
        
        \filldraw[fill=lightgray, draw=black] (2.75, 1) rectangle (2.875,1.75);
        
        \filldraw[fill=blue!10, draw=black] (2.875, 1) rectangle (5.25,1.75);
        \node[] at (4.06, 1.375) {\circled{3}{blue!10}{black} Object Type};

        \filldraw[fill=white, draw=black] (5.25, 1) rectangle (9,1.75);
        \node[] at (7.125, 1.375) {\circled{4}{white}{black} Bounds};

        \filldraw[fill=white, draw=black] (0, .25) rectangle (9,1);
        \node[] at (4.5, 0.625) {\circled{5}{white}{black} Baseline architecture address};
    \end{tikzpicture}
    }
    \caption{Prismatic capability layout. Fields \dCOne{}--\dCFive{} have the same usage as the CHERI capability (from \Cref{fig:chericap}), except for \dCThree{} which is further subdivided into \capcolor and \gradient.}
    \label{fig:prismcap}
\end{figure}

\paragraph{Hue, \gradient and \gls{otype} encoding.}
The layout of a prismatic capability is shown in \Cref{fig:prismcap}. Compared to the CHERI capability layout in \Cref{fig:chericap}, a prismatic capability subdivides the \gls{otype} into two sub-ranges reserved for enclave use.
The reserved sub-range of the \gls{otype} space is defined via \texttt{TEE\_OTYPE\_BASE}. All \glspl{otype} values defined below this are either \gradients or \gls{cheri} reserved types, whereas \glspl{otype} at or above it encode enclave instance identity in \circled{3b}{white}{black} and its \gls{epvt} metadata in \circled{3a}{white}{black}. Specifically, each prismatic
capability \gls{otype} is defined as \texttt{otype} $=$ (\texttt{\capcolor} $\ll$ \texttt{pvt\_shift}) $\mathbin{|}$ \texttt{\gradient}, where:
\begin{itemize}[nosep,leftmargin=*]
    \item \texttt{\capcolor} is an \texttt{id\_width}-bit enclave instance identifier that selects the \gls{cam} entry and \caphashtable row;
    \item \texttt{\gradient} is the remaining low-bit index into its \gls{epvt};
    \item \texttt{pvt\_shift} splits the 21 \gls{otype} bits between \texttt{\capcolor} and \texttt{\gradient} bits (e.g., \texttt{pvt\_shift} = 21 - \texttt{id\_width}).
\end{itemize}

The value of \texttt{pvt\_shift} is determined at run time from a machine-mode \gls{csr}. Making this split software-configurable preserves backward compatibility across different enclave-ID allocations without modifying the hardware datapath.   The default 6-bit \texttt{id\_width} admits up to 63 concurrent
enclaves (\capcolor$\in [1, 63]$) and a 15-bit \gradient per enclave instance. This split can be retuned without datapath changes.
The four reserved CHERI \glspl{otype}
fall outside \texttt{TEE\_OTYPE\_BASE} and are exempt from the \gls{cam} check; the
\gls{cam} gates only capabilities whose \gls{otype} lies in the enclave range. 

\paragraph{Machine-mode monopoly on \capcolor assignment.}  The \capcolor is an unforgeable, M-mode-rooted credential that untrusted code may hold and pass but never conceive. The \acron hardware
restricts creation of enclave-range \glspl{otype} to M-mode: any attempt by S- or U-mode software to seal a capability with a \capcolor it does not already hold is rejected.
Untrusted software thus cannot forge an enclave's \capcolor or mint a new one to alias it. The only source of valid enclave-owned \enclavecaps is the \gls{hmm}'s creation routine.

\subsection{Hue Management Monitor}\label{sec:tcb}
The \gls{hmm} is a minimal M-mode security monitor. It is the only software
component that can program \gls{cam} entries, access enclave memory (via M-mode exemption), and manage \capcolor and \gls{otype} encoding.

\paragraph{Boot}. The boot process atop \acron occurs in two stages. The \gls{hmm} is the first code to execute at reset. It (1) configures the \gls{cam} so that supervisor-mode software cannot access the \tcb's text region, (2) reserves one \capcolor for its own exclusive use, and (3) initializes the \caphashtable and M-mode \gls{csr} defining \texttt{pvt\_shift}. In the second stage, the \gls{hmm} loads the CheriBSD kernel and transfers control to its entry point in supervisor mode.
Because isolation between \gls{hmm} and \gls{os} is configured in the first stage, the \gls{cam} blocks OS access to \gls{hmm} memory. The OS therefore serves \gls{tee}-related requests from the application through \texttt{/dev/tee} device node (discussed further in \Cref{sec:kernel}). 

\paragraph{Creation.}
On an enclave creation request, the \gls{hmm}: 
(1) assigns a \capcolor from a bitmap allocator,
(2) measures the enclave code and static data pages in place by computing SHA-256 over their contents
(using relative virtual addresses so offline verification is independent of
load address),
(3) programs the \gls{cam} entry (with \texttt{valid=1, active=0}),
and (4) seals code and data capabilities with the assigned \capcolor.
The sealed capabilities are returned to the kernel via a shared buffer. 
The \gls{hmm} also processes \gls{caprelocs} on behalf of the
enclave, deriving a corresponding capability for each entry:
\begin{itemize}[nosep,leftmargin=*]
    \item \emph{function entries} receive enclave-wide bounds (so that \gls{pcc}-relative addressing remains valid from any function) and execute permission, derived from M-mode \gls{pcc};
    \item \emph{data entries} receive per-object bounds derived from M-mode \gls{ddc}, with constant entries further restricted to read-only.
\end{itemize}
Each resulting capability is associated with the enclave's \capcolor and written into enclave memory where indicated by the corresponding \gls{caprelocs} entry.
This ensures indirect references in the enclave use a valid \enclavecap carrying the enclave's \capcolor. 

\paragraph{Trusted interrupt handler.}
Timer interrupts arrive asynchronously and, because the untrusted \gls{os}
services them, pose a confidentiality hazard: a preempted enclave's register file must never be visible to the kernel, yet the kernel must still schedule and enforce fairness.
\acron reconciles these requirements by interposing the \gls{hmm} on the interrupt path.
When a timer fires during enclave execution, the \gls{hmm} (1) takes the trap in M-mode,
(2) saves the enclave's architectural state into a per-enclave \gls{ssa} in the enclave's own \capcolor-protected data region,
(3) scrubs the general-purpose registers, and (4) converts the machine timer interrupt into a supervisor timer interrupt that the kernel handles through its ordinary path.
Execution resumes transparently on the next enclave entry.
The \gls{ssa} is laid out as follows:

\begin{center}
\footnotesize
\begin{tabular}{lll}
\toprule
\textbf{Offset} & \textbf{Field} & \textbf{Size} \\
\midrule
0x000 & \texttt{resume\_ marker} & 8 bytes \\
0x010 & \texttt{saved\_pcc} & 16 bytes (capability) \\
0x020 & \texttt{user\_ret} & 16 bytes (capability) \\
0x030 & \texttt{regs[0..31]} & 32 $\times$ 16 bytes \\
\bottomrule
\end{tabular}
\end{center}

\emph{Non-preemptible transition windows.}
Saving the enclave's architectural state is only correct when the interrupted program counter lies in enclave instructions, where all state is live in registers and the \gls{ssa} is not being modified. 
This invariant breaks while the enclave is inside \emph{its own} entry/exit trampoline: these sequences write the \gls{ssa} themselves,
so saving the state mid-sequence could corrupt it or overwrite resume context that an \gls{ocall} is still consuming.
We therefore treat this short, statically known entry/exit sequence as \emph{non-preemptible}.
The \gls{hmm} sets a conservative boundary above it and, on a timer trap, compares the faulting \gls{pcc}
against this boundary,
declining to enter the respective trampoline again and returns via \texttt{mret}.
Since the sequence is short and non-looping, the pending tick is delivered cleanly through the kernel's normal path on the next cycle.

\emph{Idempotent resume.}
On re-entry via \EInvoke the trampoline inspects \texttt{SSA.resume\_marker}: if unset, the trampoline takes the first-entry path that establishes the stack and calls directly into enclave; if set, \texttt{SSA.resume\_marker} denotes a pending interrupt resume which \PRISM handles in M-mode.
The trampoline executes a single reserved instruction that traps directly into the \gls{hmm}.
The \gls{hmm} reloads general-purpose registers from the \gls{ssa}, sets \texttt{mepcc} \gls{csr} to \texttt{saved\_pcc}, and returns with \texttt{mret} to the enclave.
Since this resume instruction sits in the non-preemptible prefix, a timer arriving just before it retires is declined and simply re-executes on the next entry, so interrupt/resume do not expose partial state.

\paragraph{Trusted fault handler.}
A synchronous fault inside an active enclave---a \gls{cam}/temporal-safety denial or any \gls{cheri} violation---raises the same hazard as an interrupt: it occurs with the \capcolor live and enclave data in registers, yet must be reported to untrusted software. \acron \emph{does not delegate} the enclave-relevant fault causes (the \gls{cheri} exception and the \gls{cam}'s load/store access faults) to supervisor mode, so they trap to the \gls{hmm}. If the faulting \gls{pcc} carries a \capcolor, the \gls{hmm} invalidates the \gls{cam} entry, scrubs the registers, and redirects to the host's return sentry with a distinguished status---as on the interrupt path, but without saving resumable state, since a fault is fatal. The host sees a defined exit, not a register leak or crash. Faults from non-enclave code carry a non-\enclavecap \gls{pcc} and are passed unchanged to the kernel.

\paragraph{Report Generation.} The \gls{hmm} constructs and signs a \texttt{REPORT} data structure
 when a host application issues the \texttt{TEE\_IOCTL\_GEN\_ATT} \texttt{ioctl} (which drives the 
\texttt{SBI\_ENCLAVE\_GEN\_ATT} call, \Cref{tab:sgxapi}). The \texttt{REPORT} data structure has the fields:
\begin{center}
\footnotesize
\begin{tabular}{ll}
\toprule
\textbf{Field} & \textbf{Size} \\
\midrule
\texttt{chal} & \texttt{\_\!\_HASH\_SIZE\_\!\_} \\
\texttt{\capcolor} & 8 bytes \\
\texttt{encl\_measurement} &  \texttt{\_\!\_HASH\_SIZE\_\!\_} \\
\texttt{hmm\_measurement} &  \texttt{\_\!\_HASH\_SIZE\_\!\_} \\
\texttt{signature} &  \texttt{\_\!\_SIG\_SIZE\_\!\_} \\
\bottomrule
\end{tabular}
\end{center}
The host application obtains \texttt{chal}, the remote attestation challenge, over the network and passes it to the \gls{hmm} call along with \texttt{\capcolor}, the \capcolor of the enclave being reported on.
The remaining fields come from \gls{hmm} private memory: \texttt{encl\_measurement} is read from the \caphashtable entry for that \capcolor, \texttt{hmm\_measurement} is a static measurement of the \gls{hmm}'s own initial state, and \texttt{signature} is the digital signature computed over the other four fields.
\texttt{\capcolor} is 8 bytes; while the remaining fields are sized according to the hash function and digital signature algorithm installed in the \gls{hmm}.
Our prototype uses SHA256 and ED25519, so \texttt{\_\!\_HASH\_SIZE\_\!\_} is 32 bytes while \texttt{\_\!\_SIG\_SIZE\_\!\_} is 64 bytes. The \gls{hmm} uses hash and signature algorithms from the formally verified HACL$^{*}$ library~\cite{hacl}.

\paragraph{Destruction.} On a destroy request for a given \capcolor, the \gls{hmm} decodes and validates the enclave id from its \capcolor, then zeros the enclave's physical pages (data region, \gls{epvt}, and stack), along with the \gls{ssa},
so that no later occupant of the same pages or \capcolor can observe stale data or resume context.
Only after scrubbing does the \gls{hmm} invalidates the \gls{cam} entry, restoring the supervisor mode access  and letting the kernel reclaim the pages.
Since the scrub runs under the \gls{hmm}'s M-mode \gls{cam} exemption, the enclave pages are never exposed to supervisor mode in an unclean state. 
Finally, it retires the \capcolor, preventing its reuse until a full-memory sweep by the \gls{hmm} (\Cref{apdx:discussion}) revokes any \enclavecaps which have escaped the enclave memory.

\subsection{CheriBSD Support}\label{sec:kernel}

We implement a \texttt{/dev/tee} device in CheriBSD whose \texttt{ioctl}s cover
the enclave lifecycle: creation, destruction, and attestation-report
generation (see ~\Cref{tab:sgxapi}). The kernel's role is minimal: allocate memory, invoke the
\gls{hmm}, and relay \enclavecaps between the \gls{hmm} and userspace. The
kernel is explicitly untrusted: it can neither access enclave memory nor forge
\enclavecaps.

\paragraph{Memory allocation \& code transfer.}
Because a single \gls{cam} entry covers one naturally-aligned power-of-two region (\Cref{sec:cam-hw}), the kernel backs each enclave with exactly such a region.
Enclave creation begins with a single \texttt{TEE\_IOCTL\_CREATE} request, carrying a caller-supplied struct that contains:
(1) the entry virtual address and size; 
(2) the relocation metadata the \gls{hmm} requires; 
(3) the \gls{epvt} size.
After validating that the region size is page-aligned and within the maximum size limit (e.g., 64 MB in our prototype), the kernel
allocates physically contiguous pages with
power-of-two size and natural alignment. For an enclave requesting $n$ pages it
allocates $2^{\lceil\log_2 n\rceil}$ pages (at most a $2\times$ overhead).
Any extra pages are zeroed and protected by the same \gls{cam} entry.

The kernel also stages the enclave image:
it wires the enclave's source pages, copies their contents through the kernel's direct-mapped window, and rewrites the page-table entries at the enclave's existing virtual addresses with the wired mappings.
This leaves the virtual address unchanged while its physical backing becomes the region the \gls{hmm} will isolate.
Because staging precedes isolation, the region remains ordinary userspace memory until the \gls{hmm} activates the \gls{cam} entry during the creation call (\Cref{sec:tcb}), after which the \gls{cam} denies access lacking the correct \capcolor.

\paragraph{The \texttt{cap\_buff} protocol.}
The kernel communicates with the \gls{hmm} through a fixed-layout shared buffer (\texttt{cap\_buffer}), passed in via an \gls{sbi} call.
The kernel fills the buffer's input region with each enclave page's physical address and relocation metadata \gls{hmm} needs to seal the capability table (the \gls{caprelocs} offset and size, the ELF text virtual address, and the static section size).
The \gls{hmm} operates
in place on this buffer, and the kernel then copies the results into the \texttt{ioctl} output, along with the allocated \capcolor and confirmed entry address.
To the kernel, these outputs are opaque bytes it relays but cannot modify or forge.

\paragraph{Destruction.} On an application's request, the kernel triggers the destruction sequence of \Cref{sec:tcb}.
Since each open \texttt{/dev/tee} admits a single enclave, closing the device tears that enclave down. 
On return, the kernel removes the enclave's page-table mappings and frees the contiguous buffer. 

\paragraph{Attestation.} A separate \texttt{ioctl} request handles attestation-report generation (\Cref{sec:tcb}).
Here, the kernel is confined to data transport: it allocates a shared buffer, copies the caller's challenge and the target enclave's \capcolor, and invokes the \gls{hmm}.
On return, it copies the \gls{hmm}-generated report into the \texttt{ioctl} result and frees the buffer.
The kernel cannot access the private key used to construct the report, and tampering in transit is detectable by inspecting the report signature. 

\subsection{Supporting Unmodified \gls{sgx} Applications}\label{sec:sgx-apps}

\begin{table}[t]\centering\footnotesize
\begin{tabular}{lll}
\toprule
\textbf{Intel \gls{sgx}} & \textbf{Host action} & \textbf{\tcb call}\\
\midrule
\texttt{sgx\_create\_enclave}  & \texttt{TEE\_IOCTL\_CREATE}  & \texttt{ENCLAVE\_CREATE}\\
\gls{ecall}                    & \texttt{\EInvoke}             & ---\\
\gls{ocall}                    & \texttt{\EExit}& ---\\
remote attestation             & \texttt{TEE\_IOCTL\_GEN\_ATT}               & \texttt{ENCLAVE\_GEN\_ATT}\\
\texttt{sgx\_destroy\_enclave} & \texttt{close}                              & \texttt{ENCLAVE\_DESTROY}\\
\bottomrule
\end{tabular}
\caption{The \gls{sgx} life cycle on \acron.}
\label{tab:sgxapi}
\end{table}

\begin{figure}[t]
    \centering
    \includegraphics[width=\columnwidth]{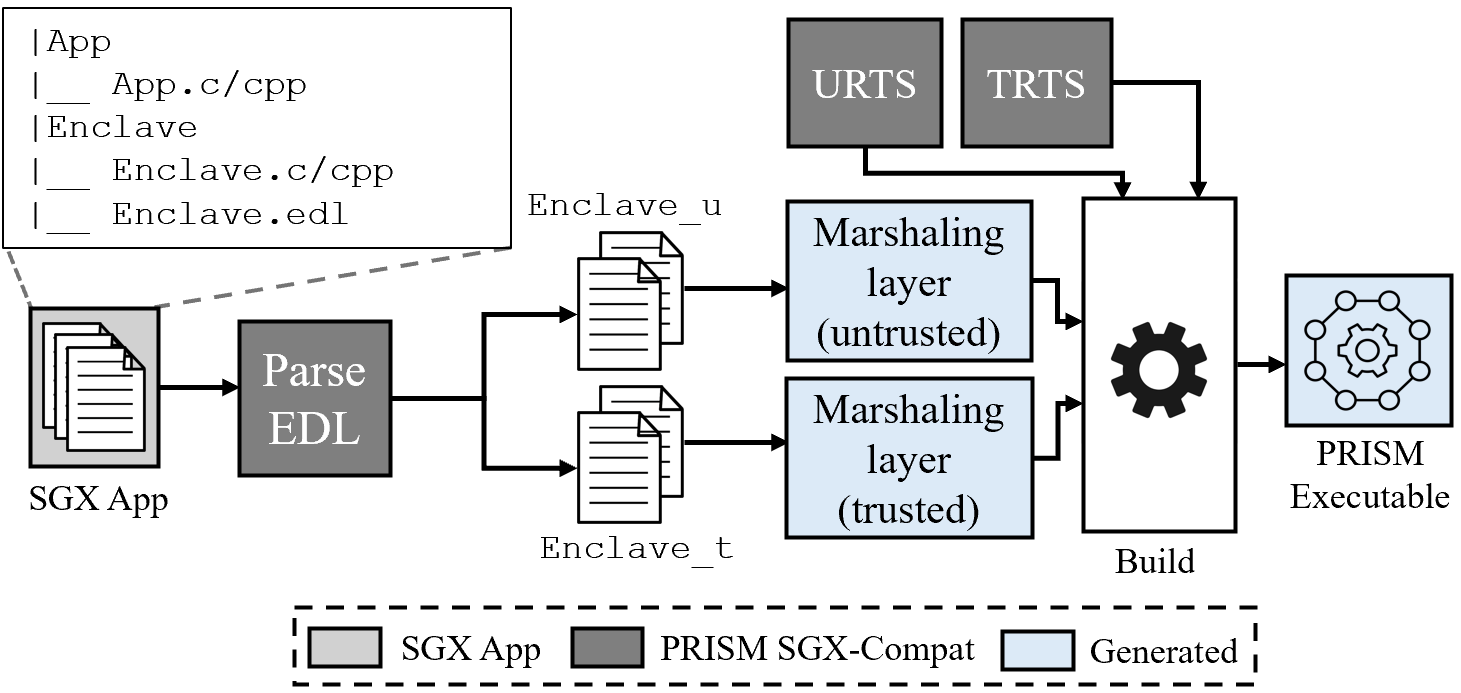}
    \caption{Processing of an Intel \gls{sgx} application with \acron's \gls{sgx}-compatibility library}
    \label{fig:sgx-compat}
\end{figure}

To support unmodified \gls{sgx} applications, we provide an \gls{sgx}-compatibility library that operates over the standard \gls{sgx} application file structure, as is depicted in Figure~\ref{fig:sgx-compat}. The resulting host/enclave interface is small (\Cref{tab:sgxapi}): the whole enclave life cycle is 
two \texttt{/dev/tee} \texttt{ioctl}s plus \texttt{close}, and enclave entry and exit are unprivileged hardware instructions that never trap to the \tcb. 
Intel SGX applications contain two source directories: 
\begin{itemize}[nosep, leftmargin=*]
    \item \texttt{App}: contains the host-side application, which creates and invokes an enclave.
    \item \texttt{Enclave}: contains the enclave code itself and an \gls{edl} file declaring the functions comprising the host-enclave interface. 
\end{itemize}
The \gls{edl} file declares a set of \glspl{ecall} and \glspl{ocall}, each annotated with a parameter direction (e.g., \texttt{[in]}, \texttt{[out]}, or \texttt{[in-out]}) and parameter semantics (e.g., \texttt{[string]} for string variables or \texttt{[count=N]} for expected buffer sizes). 

We first parse the \gls{edl} to identify the \gls{ecall} and \gls{ocall} functions, signatures, and shared data structs, producing four output files: \texttt{Enclave\_u.h}, \texttt{Enclave\_u.c/cpp}, \texttt{Enclave\_t.h}, \texttt{Enclave\_t.c/cpp}.
These files implement the \emph{untrusted} (\texttt{\_u}) and \emph{trusted} (\texttt{\_t}) \emph{marshaling layers}.
Each layer defines identical per-\gls{ecall} structs describing the packed argument layout, ensuring the host and enclave agree on the data representation across the trust boundary.
The untrusted layer implements \gls{ecall} wrapper functions and an \gls{ocall} dispatcher.
The trusted layer implements \gls{ocall} stubs and an \gls{ecall} dispatcher that reads the \gls{ecall} ID, allocates input buffers, and invokes the target function.
On return, it copies results back before freeing the temporary buffers.

At build time, the marshaling layers are combined with our untrusted and trusted runtime systems (\gls{urts} and \gls{trts}).
The \gls{trts} is linked directly into the enclave source code and supplies
(1) an \gls{ecall} trampoline for handling \gls{ecall} entry, \gls{ocall} exit, and interrupt resumption;
(2) a freestanding bounded heap allocator--adapted from dlmalloc;
(3) a linker script; and
(4) a freestanding C/math/C++ runtime---objects extracted from CheriBSD kernel. 
The \gls{urts} is linked into the host application and implements enclave management (creation/destruction), the \gls{ecall}/\gls{ocall} dispatch loop, and an assembly wrapper around \EInvoke for entry/re-entry. 
The \gls{trts}'s allocator additionally enforces temporal-safety, associating each allocation with a fresh \gradient and triggering a revocation sweep on \gradient exhaustion (\Cref{sec:temporal-safety}).

\paragraph{Outside calls (OCALLs).}
An \gls{ocall} reaches a host-side service (file I/O, network access, or memory allocation beyond the enclave's static heap) without involving the \gls{hmm}:
the enclave exits and re-enters through the normal invocation path.
The enclave writes the \gls{ocall} request to a shared host-side memory buffer.
It then saves its callee-saved registers and return address into a dedicated \gls{ocall} slot within the \gls{ssa} (\Cref{sec:tcb}, disjoint from the interrupt save area), writes an \gls{ocall}-resume marker value, and executes \EExit.
The \gls{urts} detects the \gls{ocall} via a distinguished return value, dispatches the requested function, writes the result
back into the shared buffer, and re-invokes the enclave via \EInvoke.
On re-entry the trampoline detects the \gls{ocall}-resume marker value, restores the saved registers, and returns control after the \gls{ocall} call site, as if the host function had returned normally.
The \gls{ssa} preserves continuity across this boundary, and the \gls{cam} keeps enclave memory locked while host code executes.
The shared buffer is an untrusted communication channel; the enclave must validate all data returned through it.

\subsection{Spatial and Temporal Safety inside TEE}\label{sec:temporal-safety}
CHERI provides spatial safety via its default \glsdesc{purecap} mode. Temporal safety uses the \gls{epvt} (\Cref{sec:highlevel}): a one-\gls{pvb}-per-slot bitmap at the top of the enclave's \capcolor-protected region, unreachable from outside. 
Each tracked allocation carries a \texttt{\gradient}, and each \gls{cam} entry holds the metadata to locate the \gls{epvt}. At creation, the \gls{hmm} reserves a fixed 4KB \gls{epvt} at the top of the enclave's \capcolor-protected region.

\paragraph{Allocation and enforcement.}
The trusted allocator assigns a \texttt{\gradient} for every allocation; \texttt{free} marks the slot invalid.
Since slots are never reused while fresh \gradients remain, a dangling capability can never match a later allocation, and use-after-frees fault immediately. Since validity bits are cached, revoking a slot is followed by a fence for the hardware check to observe it.

\paragraph{Batch revocation.}
On \gradient exhaustion, the allocator runs one revocation sweep: it scans the enclave's heap, globals, and live stack, clears the tag of every stored capability whose slot is revoked, and returns those slots for future allocations.
The sweep runs inside the enclave's trusted runtime---the \gls{hmm} is uninvolved, and unlike per-process CHERI revocation (or system-wide \capcolor revocation), only enclave memory is scanned.

\subsection{Remote Attestation Protocol}\label{sec:ra}

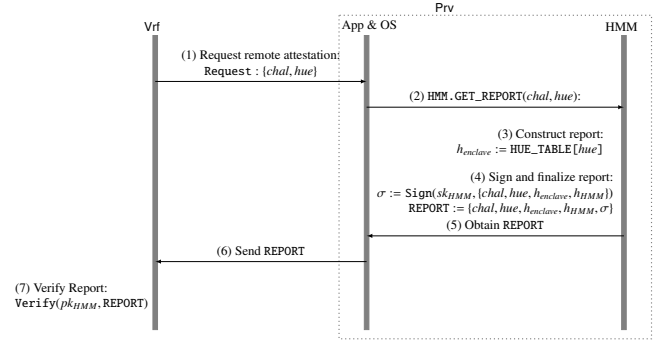
\begin{figure}[t]
\centering
\resizebox{\columnwidth}{!}{
\begin{tikzpicture}[>=latex,scale=0.9, every node/.style={scale=0.9}]
\def\ytop{0}
\def\ybottom{-8}
\def\xvrf{0.25}
\def\xapp{6}
\def\xbbl{13}
\def\y{0}
\def\dy{-0.7}

\filldraw[fill=gray, draw=gray] (\xvrf-.0625,\ytop) rectangle (\xvrf+.0625,\ybottom);
\filldraw[fill=gray, draw=gray] (\xapp-.0625,\ytop) rectangle (\xapp+.0625,\ybottom);
\filldraw[fill=gray, draw=gray] (\xbbl-.0625,\ytop) rectangle (\xbbl+.0625,\ybottom);
\filldraw[fill=none, draw=gray, dotted, thick] (\xapp-0.75,\ytop+0.55) rectangle (\xbbl+0.75,\ybottom-0.25);
\node[left] at (\xvrf+.25,\ytop+.25) {\vrf};
\node[left] at (\xapp+2.5,\ytop+.75) {\prv};
\node[left] at (\xbbl+.5,\ytop+.25) {HMM};
\node[left] at (\xapp+.95,\ytop+.25) {App \& OS};
\draw[->] (\xvrf,\y-1.25) -- (\xapp,\y-1.25) node[midway, above] {
\shortstack{(1) Request remote attestation: \\ $\texttt{Request}: \{chal, \capcolor\}$}};
\pgfmathsetmacro\y{\y+\dy}
\draw[->] (\xapp,\y-1.25) -- (\xbbl, \y-1.25) node[midway, above] {
\shortstack{(2) \texttt{HMM.GET\_REPORT}($chal, \capcolor$):}};
\pgfmathsetmacro\y{\y+\dy}
\node[align=right] at (\xbbl-2.5,\y-1.5)
{\shortstack[r]{(3) Construct report: \\
$h_{enclave} := \texttt{HUE\_TABLE[}\capcolor\texttt{]}$
}
};
\pgfmathsetmacro\y{\y+\dy+\dy}
\node[align=right] at (\xbbl-3.5,\y-1.5)
{\shortstack[r]{(4) Sign and finalize report:} \\
$\sigma := \texttt{Sign}(sk_{HMM}, \{chal, \capcolor, h_{enclave}, h_{HMM}\})$ \\
\texttt{REPORT} := $\{chal, \capcolor, h_{enclave}, h_{HMM}, \sigma\}$
};
\pgfmathsetmacro\y{\y+\dy+\dy}
\draw[->] (\xbbl,\y-1.25) -- (\xapp, \y-1.25) node[midway, above] {
\shortstack{(5) Obtain \texttt{REPORT} }};
\pgfmathsetmacro\y{\y+\dy}
\draw[->] (\xapp,\y-1.25) -- (\xvrf, \y-1.25) node[midway, above] {
\shortstack{(6) Send \texttt{REPORT} }};
\pgfmathsetmacro\y{\y+\dy}
\node[align=left] at (\xvrf-2,\y-1.5)
{\shortstack[r]{(7) Verify Report:} \\
$\texttt{Verify}(pk_{HMM}, \texttt{REPORT})$
};
\pgfmathsetmacro\y{\y+\dy+\dy}
\end{tikzpicture}
}
\caption{Remote attestation protocol for a running enclave that was assigned \capcolor by \gls{hmm}.}
\label{fig:protocol}
\end{figure}

\Cref{fig:protocol} depicts the remote attestation protocol between \vrf and \prv.
This protocol assumes the following initial state: 
\begin{inparaenum}[a)]
    \item an enclave has been created and assigned a \capcolor;
    \item the measurement of its code/data ($h_{enclave}$) has been stored in the \caphashtable;
    \item \prv has a signing key pair ($pk_{HMM}, sk_{HMM}$), for which the verification key is known by \vrf;
    \item measurement of \gls{hmm} initial code/data ($h_{HMM})$ is in \gls{hmm} storage. 
\end{inparaenum}

Remote attestation begins in (1) when \vrf sends a request containing $chal$ and the $\capcolor$ of the enclave that should be reported. The request is received by the host application. To generate a report, the host application makes an \tcb call in (2) to \texttt{HMM.GET\_REPORT}, passing the request data as inputs. 
In (3), the \gls{hmm} assembles the report fields specified in \Cref{sec:tcb}: it reads$h_{enclave}$ from the \caphashtable by indexing with the $\capcolor$, and fetches $h_{HMM}$ from its own memory. In (4), $\{chal, \capcolor, h_{enclave}, h_{HMM}\}$ are signed using $sk_{HMM}$ to obtain a signature ($\sigma$), finalizing \texttt{REPORT} as $\{chal, \capcolor, h_{enclave}, h_{HMM}, \sigma\}$.
As a result, $\sigma$ authenticates that this report was generated by \prv's \gls{hmm} at the time of the request. 
In (5), \gls{hmm} returns \texttt{REPORT} to the host application, which forwards it to \vrf in (6).
On receipt in (7), \vrf verifies \texttt{REPORT} by:
\begin{inparaenum}[i)]
    \item unpacking it into \{$chal$, $\capcolor$, $h_{enclave}$, $h_{\tcb}, \sigma\}$; 
    \item validating $\sigma$ against $pk_{\tcb}$;
    \item checking that $chal, \capcolor$ match the original request;
    \item checking $h_{enclave}$ against the expected code/data pages for the enclave;
    \item checking $h_{\tcb}$ against the expected configuration.
\end{inparaenum}

\section{Evaluation}\label{sec:eval}
We evaluate \acron across three dimensions: performance, security, and hardware area cost.
For performance, we use the BESSPIN-GFE security evaluation platform, which supports the CHERI-Toooba softcore out-of-the-box.
We replace the standard core with our \acron CHERI-Toooba, an extension of the CHERI-RISC-V Toooba \gls{fpga} softcore (RV64ACDFIMSUxCHERI) built on the open-source Bluespec RISC-V 64-bit Toooba core. 
We synthesize the \gls{soc} at 25MHz (default for BESSPIN-GFE) targeting the Xilinx Virtex UltraScale+ VCU118 FPGA. 
We answer the following research questions:
\begin{enumerate}[label=\textbf{RQ\arabic*},labelsep*=0pt]
    \item:~{How does \acron impact the baseline processor in terms of performance compared to state-of-the-art capability-based temporal-safety mechanisms (e.g., \PICASSO~\cite{Gulmez26})?}\label{rq:performance-vs-picasso}
    \item:~{How do overheads introduced by \acron for enclave execution compare to overhead introduced by Intel SGX?}\label{rq:performance-vs-sgx}
    \item:~{How does \acron impact the baseline processor in terms of fixed hardware, energy, and memory cost?}\label{rq:fixed-costs}
    \item:~{How does \acron security hold against the adversary model defined in Section~\ref{sec:threatmodel}?}\label{rq:security}
\end{enumerate}

\subsection{Performance}

This section answers \ref{rq:performance-vs-picasso} and \ref{rq:performance-vs-sgx}.
For \ref{rq:performance-vs-picasso}, we run a series of test programs from SPEC CPU 2006~\cite{spec06} benchmarks and compare the performance of \acron against \PICASSO~\cite{Gulmez26}. For \ref{rq:performance-vs-sgx}, we evaluate three open-source \gls{sgx} applications and compare \acron's relative performance overhead to \gls{sgx}. 

\subsubsection{Benchmarks}

\begin{figure}[t]
    \centering
    \resizebox{\columnwidth}{!}{
\begin{tikzpicture}
    \begin{axis}[
      width=\columnwidth,
      height=\appSubPlotHeightN,
      ybar=2pt,
      bar width=7pt,
      ymin=0, ymax=1.8,
      ytick={0,0.5,1,1.5},
      ylabel={\footnotesize{Overhead Factor (Cycles)}},
      symbolic x coords={A,B,C,D,E,F,G,H,I},
      xticklabels={bzip2, gobmk, hmmer, sjeng, libquantum, h264ref, omnetpp, xalancbmk, \textbf{geomean}},
      xtick=data,
      xticklabel style={font=\footnotesize, rotate=25},
      legend style={
        at={(0.5,1.5)},
        anchor=north,
        legend columns=-1,
        font=\footnotesize,
      },
      cycle list={
        {fill=barwhite, draw=black, thick},
        {fill=barlightgray, draw=black, thick}
      },
      enlarge x limits=0.05,
      scale only axis,
      after end axis/.code={
        \foreach \tick/\v in {A/1.0, B/1.0, C/1.0, D/1.04, E/1.01, F/1.01, G/1.38, H/1.02, I/1.05} {
          \pgfmathsetmacro{\capy}{min(\v,1.5)}
          \node[font=\tiny,inner sep=1pt,anchor=south] at ([xshift=-6pt] axis cs:\tick,\capy) {\v};
        }
        \foreach \tick/\v in {A/1.0, B/1.0, C/1.0, D/1.05, E/1.09, F/0.99, G/1.37, H/1.03, I/1.06} {
          \pgfmathsetmacro{\capy}{min(\v,1.5)}
          \node[font=\tiny,inner sep=1pt,anchor=south] at ([xshift=6pt] axis cs:\tick,\capy) {\v};
        }
      },
    ]
    \addlegendentry{\PICASSO}
    \addplot coordinates {(A,1) (B,1) (C,1) (D,1.04) (E,1.01) (F,1.01) (G,1.38) (H,1.02) (I,1.05)};
    \addlegendentry{\acron}
    \addplot coordinates {(A,1) (B,1) (C,1) (D,1.05) (E,1.09) (F,0.99) (G,1.37) (H,1.03) (I,1.06)};
    \addlegendimage{line legend, thick, black, dotted}
    \addlegendentry{Baseline (1.0x)}
    \draw[thick, black, dotted] (rel axis cs:0,0.55) -- (rel axis cs:1,0.55);
    \end{axis}

  \end{tikzpicture}
  }
    \vspace{-2em}
    \caption{Comparison of overhead factor between \acron and \texttt{PICASSO}~\cite{Gulmez26} for SPEC CPU 2006~\cite{spec06} benchmarks.}
    \label{fig:spec}
\end{figure}
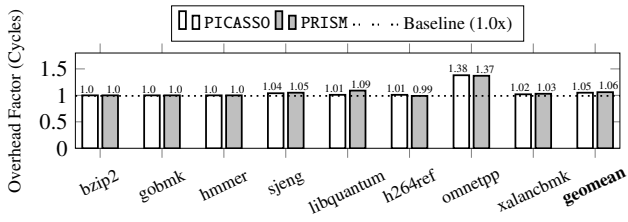

\Cref{fig:spec} shows the SPEC CPU 2006 comparison between \acron and \PICASSO.
Despite providing a stronger security guarantee (enclave execution and temporal safety vs. temporal safety alone) compared to \PICASSO, \acron incurs effectively the same overhead factor as \PICASSO across most applications: a $\approx$1.06x geomean increase for \acron compared to $\approx$1.05x for \PICASSO.
The one exception is \texttt{libquantum}, where \acron's overhead factor departs from \PICASSO's.
These results answer \ref{rq:performance-vs-picasso}: \acron imposes only modest overhead on baseline execution compared to \texttt{PICASSO}.

\subsubsection{SGX Applications}\label{sec:usecases}

We evaluate three \gls{sgx} applications:

\paragraph{sgx-kmeans}~\cite{sgx_kmeans_repo} runs k-means clustering inside the enclave on points supplied by the host. Using the repository's generation scripts, we produce input files with sizes ranging from 100K to 1M points in 100K increments.
To account for variance across cluster sizes, we average results for each input file while varying the number of clusters ($K=2$ to $8$). 

\paragraph{SGX\_SQLite}~\cite{sgx_sqlite_repo} hosts a SQL database and runs queries in an enclave. We use a subset of the \texttt{sqllogictests} suite~\cite{sqllogictest}, specifically the \texttt{select} and \texttt{index\_between} (\texttt{ib}) cases. 

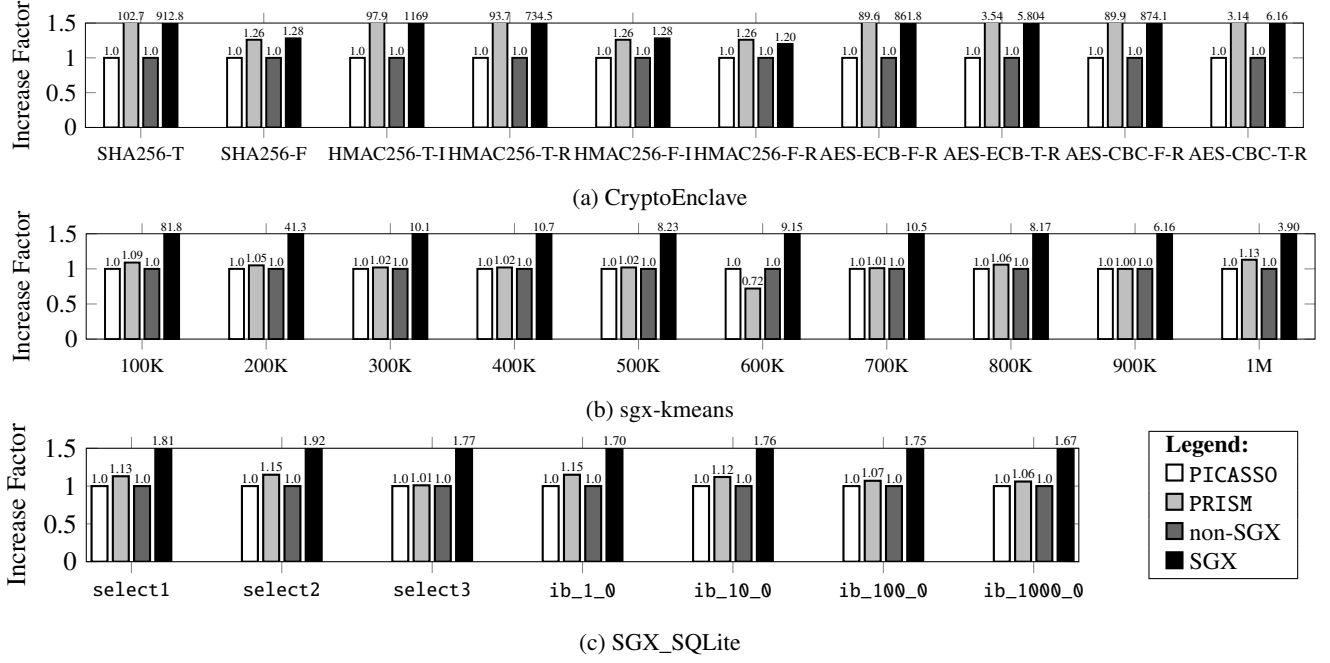
\begin{figure*}[t]
    \begin{subfigure}{\textwidth}
  \centering
  \resizebox{\textwidth}{!}{
  \begin{tikzpicture}
    \begin{axis}[
      width=\textwidth,
      height=\appSubPlotHeightN,
      ybar=2pt,
      bar width=6pt,
      ymin=0, ymax=1.5,
      ytick={0,0.5,1,1.5},
      ylabel={Increase Factor},
      symbolic x coords={A,B,C,D,E,F,G,H,I,J},
      xticklabels={SHA256-T, SHA256-F, HMAC256-T-I, HMAC256-T-R, HMAC256-F-I, HMAC256-F-R, AES-ECB-F-R, AES-ECB-T-R, AES-CBC-F-R, AES-CBC-T-R},
      xtick=data,
      xticklabel style={font=\footnotesize},
      cycle list={
        {fill=barwhite, draw=black, thick},
        {fill=barlightgray, draw=black, thick},
        {fill=bardarkgray, draw=black, thick},
        {fill=barblack, draw=black, thick},
      },
      enlarge x limits=0.05,
      scale only axis,
      after end axis/.code={
        \foreach \tick/\v in {A/1.0, B/1.0, C/1.0, D/1.0, E/1.0, F/1.0, G/1.0, H/1.0, I/1.0, J/1.0} {
          \pgfmathsetmacro{\capy}{min(\v,1.5)}
          \node[font=\tiny,inner sep=1pt,anchor=south] at ([xshift=-12pt] axis cs:\tick,\capy) {\v};
        }
        \foreach \tick/\v in {A/102.7, B/1.26, C/97.9, D/93.7, E/1.26, F/1.26, G/89.6, H/3.54, I/89.9, J/3.14} {
          \pgfmathsetmacro{\capy}{min(\v,1.5)}
          \node[font=\tiny,inner sep=1pt,anchor=south] at ([xshift=-4pt] axis cs:\tick,\capy) {\v};
        }
        \foreach \tick/\v in {A/1.0, B/1.0, C/1.0, D/1.0, E/1.0, F/1.0, G/1.0, H/1.0, I/1.0, J/1.0} {
          \pgfmathsetmacro{\capy}{min(\v,1.5)}
          \node[font=\tiny,inner sep=1pt,anchor=south] at ([xshift=4pt] axis cs:\tick,\capy) {\v};
        }
        \foreach \tick/\v in {A/912.8, B/1.28, C/1169, D/734.5, E/1.28, F/1.20, G/861.8, H/5.804, I/874.1, J/6.16} {
          \pgfmathsetmacro{\capy}{min(\v,1.5)}
          \node[font=\tiny,inner sep=1pt,anchor=south] at ([xshift=12pt] axis cs:\tick,\capy) {\v};
        }
      },
    ]
    \addplot coordinates {(A,1) (B,1) (C,1) (D,1) (E,1) (F,1) (G,1) (H,1) (I,1) (J,1)};
    \addplot coordinates {(A,102.7) (B,1.26) (C,97.9) (D,93.6) (E,1.26) (F,1.26) (G,89.7) (H,3.53) (I,89.9) (J,3.14)};
    \addplot coordinates {(A,1) (B,1) (C,1) (D,1) (E,1) (F,1) (G,1) (H,1) (I,1) (J,1)};
    \addplot coordinates {(A,2.5) (B,1.28) (C,2.5) (D,2.5) (E,1.28) (F,1.20) (G,2.5) (H,5.8) (I,2.5) (J,6.16)};
    \end{axis}
  \end{tikzpicture}
  }
  \caption{CryptoEnclave}
  \vspace{-.5em}
  \label{fig:crypto-normalized}
\end{subfigure}
    \begin{subfigure}{\textwidth}
  \centering
  \resizebox{\textwidth}{!}{
  \begin{tikzpicture}
    \begin{axis}[
      width=\textwidth,
      height=\appSubPlotHeightN,
      ybar=2pt,
      bar width=6pt,
      ymin=0, ymax=1.5,
      ytick={0,0.5,1,1.5},
      ylabel={Increase Factor},
      enlarge x limits=0.1,
      symbolic x coords={A,B,C,D,E,F,G,H,I,J},
      xticklabels={100K, 200K, 300K, 400K, 500K, 600K, 700K, 800K, 900K, 1M},
      xtick=data,
      xticklabel style={font=\footnotesize},
      cycle list={
        {fill=barwhite, draw=black, thick},
        {fill=barlightgray, draw=black, thick},
        {fill=bardarkgray, draw=black, thick},
        {fill=barblack, draw=black, thick},
      },
      enlarge x limits=0.05,
      scale only axis,
      after end axis/.code={
        \foreach \tick/\v in {A/1.0, B/1.0, C/1.0, D/1.0, E/1.0, F/1.0, G/1.0, H/1.0, I/1.0, J/1.0} {
          \pgfmathsetmacro{\capy}{min(\v,1.5)}
          \node[font=\tiny,inner sep=1pt,anchor=south] at ([xshift=-12pt] axis cs:\tick,\capy) {\v};
        }
        \foreach \tick/\v in {A/1.09, B/1.05, C/1.02, D/1.02, E/1.02, F/0.72, G/1.01, H/1.06, I/1.00, J/1.13} {
          \pgfmathsetmacro{\capy}{min(\v,1.5)}
          \node[font=\tiny,inner sep=1pt,anchor=south] at ([xshift=-4pt] axis cs:\tick,\capy) {\v};
        }
        \foreach \tick/\v in {A/1.0, B/1.0, C/1.0, D/1.0, E/1.0, F/1.0, G/1.0, H/1.0, I/1.0, J/1.0} {
          \pgfmathsetmacro{\capy}{min(\v,1.5)}
          \node[font=\tiny,inner sep=1pt,anchor=south] at ([xshift=4pt] axis cs:\tick,\capy) {\v};
        }
        \foreach \tick/\v in {A/81.8, B/41.3, C/10.1, D/10.7, E/8.23, F/9.15, G/10.5, H/8.17, I/6.16, J/3.90} {
          \pgfmathsetmacro{\capy}{min(\v,1.5)}
          \node[font=\tiny,inner sep=1pt,anchor=south] at ([xshift=12pt] axis cs:\tick,\capy) {\v};
        }
      },
    ]
    \addplot coordinates {(A,1) (B,1) (C,1) (D,1) (E,1) (F,1) (G,1) (H,1) (I,1) (J,1)};
    \addplot coordinates {(A,1.09) (B,1.05) (C,1.02) (D,1.02) (E,1.02) (F,0.72) (G,1.01) (H,1.06) (I,1.00) (J,1.13)};
    \addplot coordinates {(A,1) (B,1) (C,1) (D,1) (E,1) (F,1) (G,1) (H,1) (I,1) (J,1)};
    \addplot coordinates {(A,81.8) (B,41.3) (C,10.1) (D,10.7) (E,8.23) (F,9.15) (G,10.5) (H,8.17) (I,6.16) (J,3.90)};
    \end{axis}
  \end{tikzpicture}
  }
  \caption{sgx-kmeans}
  \vspace{-2.5em}
  \label{fig:kmeans-normalized}
\end{subfigure}
    \begin{subfigure}{\textwidth}
  \centering
  \begin{tikzpicture}
    \begin{axis}[
      width=0.75\textwidth,
      height=\appSubPlotHeightN,
      ybar=2pt,
      bar width=6pt,
      ymin=0, ymax=1.5,
      ytick={0,0.5,1,1.5},
      ylabel={Increase Factor},
      symbolic x coords={A,B,C,D,E,F,G},
      xticklabels={\texttt{select1}, \texttt{select2}, \texttt{select3}, \texttt{ib\_1\_0}, \texttt{ib\_10\_0}, \texttt{ib\_100\_0}, \texttt{ib\_1000\_0}},
      xtick=data,
      xticklabel style={font=\footnotesize},
      cycle list={
        {fill=barwhite, draw=black, thick},
        {fill=barlightgray, draw=black, thick},
        {fill=bardarkgray, draw=black, thick},
        {fill=barblack, draw=black, thick},
      },
      enlarge x limits=0.05,
      scale only axis,
      after end axis/.code={
        \foreach \tick/\v in {A/1.0, B/1.0, C/1.0, D/1.0, E/1.0, F/1.0, G/1.0} {
          \pgfmathsetmacro{\capy}{min(\v,1.5)}
          \node[font=\tiny,inner sep=1pt,anchor=south] at ([xshift=-12pt] axis cs:\tick,\capy) {\v};
        }
        \foreach \tick/\v in {A/1.13, B/1.15, C/1.01, D/1.15, E/1.12, F/1.07, G/1.06} {
          \pgfmathsetmacro{\capy}{min(\v,1.5)}
          \node[font=\tiny,inner sep=1pt,anchor=south] at ([xshift=-4pt] axis cs:\tick,\capy) {\v};
        }
        \foreach \tick/\v in {A/1.0, B/1.0, C/1.0, D/1.0, E/1.0, F/1.0, G/1.0} {
          \pgfmathsetmacro{\capy}{min(\v,1.5)}
          \node[font=\tiny,inner sep=1pt,anchor=south] at ([xshift=4pt] axis cs:\tick,\capy) {\v};
        }
        \foreach \tick/\v in {A/1.81, B/1.92, C/1.77, D/1.70, E/1.76, F/1.75, G/1.67} {
          \pgfmathsetmacro{\capy}{min(\v,1.5)}
          \node[font=\tiny,inner sep=1pt,anchor=south] at ([xshift=12pt] axis cs:\tick,\capy) {\v};
        }
      },
    ]
    \addplot coordinates {(A,1) (B,1) (C,1) (D,1) (E,1) (F,1) (G,1)};
    \addplot coordinates {(A,1.13) (B,1.15) (C,1.01) (D,1.15) (E,1.12) (F,1.07) (G,1.06)};
    \addplot coordinates {(A,1) (B,1) (C,1) (D,1) (E,1) (F,1) (G,1)};
    \addplot coordinates {(A,1.81) (B,1.92) (C,1.77) (D,1.70) (E,1.76) (F,1.75) (G,1.67)};
    \end{axis}
     \node[anchor=west, inner sep=26pt, font=\small] at (current axis.east) {
        \begin{tabular}{{|l|}}
          \hline
          \textbf{Legend:}\\
          \tikz{\fill[barwhite,draw=black,thick] (0,0) rectangle (6pt,6pt);} \PICASSO\\
          \tikz{\fill[barlightgray,draw=black,thick] (0,0) rectangle (6pt,6pt);} \acron\\
          \tikz{\fill[bardarkgray,draw=black,thick] (0,0) rectangle (6pt,6pt);} non-SGX\\
          \tikz{\fill[barblack,draw=black,thick] (0,0) rectangle (6pt,6pt);} SGX\\
          \hline
        \end{tabular}
      };
  \end{tikzpicture}
  \vspace{-2.5em}
  \caption{SGX\_SQLite}
  \vspace{-1.5em}
  \label{fig:sqlite-normalized}
\end{subfigure}
    \caption{Normalized run-time comparison: for each application, we compare the end-to-end run-time increase of \texttt{PICASSO}$\rightarrow$PRISM to non-SGX$\rightarrow$SGX. \texttt{PICASSO} and non-SGX execution serve as the two relative baselines. The reported increase factors for \acron are relative to \texttt{PICASSO} and for SGX are relative to non-SGX execution.}
    \label{fig:normalized-run-time}
\end{figure*}

\paragraph{CryptoEnclave}~\cite{crypto_enclave_repo} is an in-enclave cryptographic library supporting SHA256, HMAC-SHA256, and AES. The host application selects an algorithm and the input message, and the enclave calculates and returns the result. Our evaluation set is the repository's own test cases, which vary the cryptographic function, plaintext input type, and key input type.
To achieve a fair comparison despite the clock-speed difference between the x86 \gls{cpu} and the Toooba \gls{fpga} softcore, we normalize \acron run time to \PICASSO and \gls{sgx} enclave run time to non-\gls{sgx} (x86) enclave execution, then compare the relative increases.
\Cref{fig:normalized-run-time} shows the result. Since \acron does not support memory encryption, we configure tested \gls{sgx} enclaves in ``pre-release'' mode to exclude memory encryption.
For \texttt{CryptoEnclave} (\ref{fig:crypto-normalized}), the input-text variants (T) show the highest relative increase as these operations have the shortest pure execution time and are sensitive to fixed enclave life-cycle costs (\Cref{sec:fixed-costs}). On all file-taking operations (F), \acron matches or improves on \gls{sgx} relative to its baseline.

\acron's advantage is most evident in \texttt{SGX\_SQLite} and \texttt{sgx-kmeans}. For \texttt{SGX\_SQLite} (\Cref{fig:sqlite-normalized}), the increase under \acron is 1.06x--1.15x, compared to 1.67x--1.81x for SGX on the same test cases.
For \texttt{sgx-kmeans} (\Cref{fig:kmeans-normalized}), the gap is wider: \acron ranges 0.72x--1.13x against SGX's 3.90x--81.8x.

We attribute \acron's advantage to two factors: First, \acron's lightweight domain transition: enclave entry and exits execute as a single unprivileged instruction handled entirely by the \gls{cam} in hardware, with no trap to the \gls{hmm}. \gls{sgx}, in contrast, mediates every EENTER/EEXIT through microcode and page-level EPCM validation~\cite{costan2016intel}.
Second, \acron's preemption path avoids SGX's Asynchronous Enclave Exit mechanism, which saves and later restores large \gls{cpu} state and restores it via the ERESUME path~\cite{costan2016intel}. \acron instead reuses the same fast entry logic for both initial entry and resumption. Since \texttt{sgx-kmeans} and \texttt{SGX\_SQLite} issue frequent ECALLs/OCALLs and memory accesses relative to their compute time, these per-transition savings compound into a smaller end-to-end increase than SGX incurs.

These results answer \textbf{RQ2}: \acron consistently incurs substantially lower overhead than Intel \gls{sgx} across all evaluated applications, owing to its hardware-mediated, trap-free domain transitions in place of SGX's microcode-mediated entry/exit and exception-handling.
Appendix~\ref{apdx:prism-picasso-sgx} extends the analysis in \Cref{sec:usecases} of the \PICASSO--\acron difference. 

\subsection{Fixed costs}\label{sec:fixed-costs}

To answer \ref{rq:fixed-costs}, we analyze \acron's impact on CHERI-Toooba processor logic, energy, and memory.
\paragraph{Impact on baseline CHERI-Toooba.} We synthesize \acron CHERI-Toooba for the VCU118 \gls{fpga}. \Cref{tab:hwcost} shows power and resource utilization reported by Xilinx Vivado 2019.1. The overheads over the unmodified CHERI-Toooba are moderate: $\approx$16\% in logic, $\approx$11\% in registers, $\approx$8\% in memory.

\begin{table}[t!]
    \centering
    \caption{Area costs of \acron and \PICASSO implemented on the CHERI-Toooba core compared to standalone CHERI-Toooba (baseline). Values are recorded by synthesizing and implementing each design on VCU118 @ 25MHz.}\label{tab:hwcost}
    \resizebox{\columnwidth}{!}{
    \Large
    \begin{tabular}{rc rcc rcc}\toprule
    & \multicolumn{1}{c}{\textbf{Baseline}}
    & \multicolumn{2}{c}{\textbf{\PICASSO}}
    & \multicolumn{2}{c}{\textbf{\acron}} \\
    & \textbf{value} & \textbf{value} & \textbf{$\Delta$ (\%)} & \textbf{value} & \textbf{$\Delta$ (\%)} \\ \midrule
    LUTs      & {688096}  & {720347}  & {32251(+4.69)}  & {795776} & {107680 (+15.6)} \\
    Memory    & {20113}   & {21611}   & {1498(+7.65)}   & {21611}  & {1498(+7.65)} \\
    Registers & {419300}  & {445129}  & {25829(+6.16)}  & {465671} & {46371(+11.00)} \\
    \bottomrule
    \end{tabular}}
\end{table}
\paragraph{\gls{hmm} Memory Footprint.}
The \gls{hmm} occupies 273.7KB. It is a modified version of the RISC-V Proxy Kernel and Boot Loader~\cite{bbl_repo}. Our changes account for 6286 lines of code, 4354 of which are due to importing supporting files from HACL*~\cite{hacl}. These changes account for a 126.6KB increase of the memory footprint.
\paragraph{Per-enclave instance operations.}
Per-instance operations, such as create and destroy, have constant cost.
In the evaluated \gls{sgx} applications this cost was consistent across test cases: enclave creation ranged from 15.63--17.27 seconds and destruction from 1.04--1.71 seconds.
Although enclave creation has a per-page operation cost (for hashing code and data to obtain the enclave's attestation measurement), its cost is dominated by the \texttt{dlopen} of the enclave shared object and the contiguous allocation operation whose cost depends on total DRAM size rather than allocation size.
Enclave destruction is dominated by the teardown of enclave page table entries, which scales with page count.

\subsection{Security Analysis}\label{sec:security}

To answer \ref{rq:security}, we revisit  \acron's desired security properties of from Section~\ref{sec:goals-and-challenges}, and argue our design and implementation upholds them.

\textit{\ref{prop:exclusiveownership} Exclusive Ownership Management.} Since \adv controls the \gls{os} and can invoke arbitrary S-/U-mode operations, \capcolor ownership must be unforgeable from that privilege level.
\acron restricts creation of enclave-range \gls{otype} values to M-mode: any S- or U-mode attempt to access a capability with a \capcolor it does not already hold is rejected, so only the \gls{hmm}'s creation routine can produce a valid enclave-owned \enclavecap.
Combined with \gls{cheri}'s monotonicity guarantee: new capabilities can only be narrowed via derivation from a valid capability, \adv cannot synthesize, upgrade, or alias a capability carrying an existing enclave's \capcolor.
\adv may hold and relay \enclavecaps opaquely (e.g., as a part of \gls{sgx}-compatibility marshaling), but cannot mint, retype, or widen them. \Capcolor ownership therefore remains rooted exclusively in the M-mode \gls{hmm}, regardless \adv's \gls{os}-level privileges.

\textit{\ref{prop:exclusivememory} Physical Memory Exclusivity}.
\adv may modify page tables and other \gls{os}-managed structures at will, so isolation cannot depend on honest virtual-to-physical translation.
\acron's \capvalidtable checks ownership on the post-translation physical address, authorizing the access by the used capability's \capcolor and denying any access whose \gls{cam} entry is inactive or whose \capcolor mismatches.
Since this check occurs post-translation, \adv can neither redirect an enclave's accesses to attacker-controlled memory through page-table manipulation nor remap pages into an active enclave's protected range.
Only M-mode is exempt; it is used solely by the \gls{hmm} to manage enclave memory during its life cycle. Since \adv cannot execute in M-mode, this exemption grants no ability to bypass the checks. 

\textit{\ref{prop:domaintransitions} Safe Domain Transitions}.
Since \adv controls OS scheduling and interrupts, domain transitions must not leak enclave state or execute under the wrong \capcolor. 
Entry and exit via \EInvoke/\EExit are unprivileged and trap-free, with \gls{cam} activation tied to the instruction commit;
on out-of-order cores, \acron suppresses speculative redirects so no instruction runs under the enclave's \capcolor before it is active.
On exit, the \gls{trts} scrubs general-purpose and capability registers, so no enclave secret in them is exposed to untrusted code. On preemption, the \gls{hmm} checkpoints the state into the \gls{cam}-protected \gls{ssa} and scrubs those registers before forwarding to the \gls{os}. 
Since the \gls{ssa} is only reachable via M-mode or a matching \capcolor, \adv learns nothing beyond an ordinary timer interrupt.
A non-preemptible trampoline boundary and idempotent M-mode resume prevent interrupt racing the enclave entry/exit from capturing inconsistent state.

\textit{\ref{prop:full-mem-safety} Spatial and Temporal Safety.} Spatial safety follows from \gls{cheri}'s guarantee that every access is mediated by an unforgeable, narrowing capability, preventing accesses outside an object's embedded bounds.
Temporal safety comes from the \gls{epvt}: each capability to the enclave's heap carries a unique identifier, which are retracted in the \gls{epvt} upon \texttt{free}, resulting in efficient detection of use-after-frees.
Since the \gls{epvt} shares the enclave's \gls{cam} entry, \adv cannot alter it.
Retraction and full capability revocation runs within the \gls{trts}, without involving untrusted code. 

\paragraph{In-enclave use-after-free detection.} We validate temporal memory safety using the heap use-after-free (CWE-416) test cases from the NIST Juliet suite~\cite{nist_juliet_2017}, executed unmodified inside a \acron enclave. All CWE-416 test cases (both C and C++ variants) are compiled into a single enclave. The untrusted host invokes each vulnerable (\texttt{bad}) and benign (\texttt{good}) function through the standard \gls{sgx} \gls{ecall} interface and records the outcome. \acron detected all 392 vulnerable cases and produced no false positives across the 392 benign cases. We discover a spatial memory safety violation in the CryptoEnclave application, which is further discussed in \Cref{apdx:discussion}.

\textit{\ref{prop:attestation} Secure Measurement and Attestation}. We validate~\ref{prop:attestation} by formally modeling the protocol in \Cref{fig:protocol} (\Cref{sec:ra}) using Tamarin Prover~\cite{tamarin}, a symbolic protocol verification tool.
Protocol steps and \adv abilities are modeled as a multiset of \emph{rules}, each rule having its own set of preconditions and postconditions (\emph{facts}). Security properties are specified as \emph{lemmas}. Given the set of rules and lemmas, the Tamarin Prover symbolically evaluates rules in arbitrary order, to determine whether all lemmas are upheld (further details in~\cite{tamarin-manual}).

We prove the following security sub-properties for \acron's attestation protocol, which together support \ref{prop:attestation}:
\begin{enumerate}[nosep,leftmargin=*]
    \item \textit{Report Authenticity}: all reports obtained by \vrf only pass verification if signed using \prv's secret key;
    \item \textit{\gls{hmm} Integrity}: the \gls{hmm} code reported by \prv matches the expected value by \vrf;
    \item \textit{Enclave Integrity}: the pages loaded into an enclave with a particular \capcolor match the expected value by \vrf;
    \item \textit{Key Secrecy}: report data sent by \prv does not allow \adv to learn secret key used to sign it;
    \item \textit{Freshness}: \vrf only accepts reports produced with a $chal$ sent as a part of their request.
\end{enumerate}
Each property maps to a lemma verified by Tamarin. Full specification of our rule-based model of \acron-based protocol and lemmas specifying these security properties are in Appendix~\ref{apdx:tamarin}.

All lemmas run in $\approx$0.45 seconds with $\approx$168 MiB peak memory usage. This uniformity reflects the shared underlying model (e.g., the protocol outlined in \Cref{fig:protocol}), and the difference in the proof steps arises from how early pruning occurs. All lemmas except for Key Secrecy require complete exploration of the model trace. In contrast, Key Secrecy is completed after two steps since it is resolved by direct absence of attacker knowledge derivations, which is determinable without complete protocol exploration.

\section{Related Work}\label{sec:rw}
To our knowledge, \acron is the first capability-based \gls{tee} to combine hardware-enforced enclave isolation, in-enclave spatial \emph{and} temporal safety, and remote attestation under an untrusted paging \gls{os}, while running unmodified \gls{sgx} applications on a real \gls{rtl} implementation. Prior work achieves only subsets of these: software techniques harden \gls{tee} code without architectural enforcement; \gls{cheri} extensions add memory safety but no enclave isolation; RISC-V \glspl{tee} isolate enclaves but leave their code's memory safety to software; and earlier capability-based \glspl{tee} forgo virtual memory, attestation, or a hardware realization.

\paragraph{Memory Safety for \glspl{tee}.} Prior work has also targeted memory safety within enclaves.
SGXBounds~\cite{Dmitrii17} extends LLVM to encode bounds into pointers and instrument accesses with bounds check.
MPTEE~\cite{zhao2020mptee} uses Intel \gls{mpx} with a cross-over memory layout to enable bounds checking across permission regions.
RustTEE~\cite{wan2020rustee} and Rust-SGX~\cite{wang2019towards} lean on Rust's memory-safety guarantees. The former provides a Rust compiler for ARM TrustZone, the latter a Rust layer over the Intel SGX SDK, with the Rust-to-C \gls{ffi} formally modeled for safety.
Unlike these compiler- and language-based approaches, \acron enforces both spatial and temporal safety architecturally. 

\paragraph{CHERI extensions}. A growing body of work extends \gls{cheri}'s baseline capability model to address security challenges beyond spatial safety.
Cornucopia~\cite{WesleyFilardo20,Filardo24} integrates a \texttt{malloc} revocation shim into CheriBSD for heap temporal safety, quarantining freed memory and sweeping only when the quarantined reaches a threshold size.
PICASSO~\cite{Gulmez26} adds \emph{colored capabilities} for provenance tracking, enabling more efficient use-after-free protection via bulk retraction.
BLACKOUT~\cite{ElAtali25} adds \emph{blinded capabilities} for added side-channel protection via taint tracking.
CapChecker~\cite{cheng2025adaptive} extends the capability model to hardware accelerators, mediating their memory interface as if they issued native capabilities.
CHERI-SIMT~\cite{naylor2026cheri} adapts capability protections to GPUs via a capability-metadata register file that exploits redundant metadata across hardware threads to cut storage overhead.
CHERI-Crypt~\cite{jackson2025cheri} adds an encryption engine to a CHERI-RISC-V32 processor for transparent memory encryption of sealed capabilities against physical attacks, introducing instruction variants that carry an encryption-required bit.

\paragraph{RISC-V TEEs.}
SANCTUM~\cite{sanctum} proposes an SGX-like design in which enclaves occupy fixed physical memory ranges, using a dual page-table lookup to keep enclave and OS page tables from referencing each other's memory.
Keystone~\cite{keystone} isolates enclaves using \gls{pmp}, with an M-mode security monitor that creates isolated enclave regions.
Timber-V~\cite{timber-v} targets embedded systems, using two-bit tagged memory to separate normal and trusted domains with MPU assistance.
CURE~\cite{cure} offers highly configurable enclaves, including exclusive enclave-to-peripheral assignment and cache-based side-channel protection.
Penglai~\cite{penglai} introduces a Guarded Page Table and Mountable Merkle Tree for scalable page-level isolation and integrity, supporting thousands of concurrent enclaves through a lightweight monitor and fast creation via forking of a special-purpose shadow enclave.
These designs typically place an M-mode monitor at highest privilege for enclave management;
Dorami~\cite{dorami} instead splits the monitor from the firmware via two isolated M-mode compartments. 
All of them isolate enclaves through page tables or \gls{pmp}, reconfigured by an M-mode monitor on every enclave switch. \acron instead checks a capability's \capcolor at access time, so entry and exit remain unprivileged and trap-free. \acron further guarantees spatial and temporal safety \emph{within} the enclave and enforces isolation on physical addresses rather than page-table configuration.

\begin{table}[t]
\centering
\resizebox{\columnwidth}{!}{
\begin{tabular}{lccc}
\toprule
\textbf{Property} & \textbf{CHERI-TrEE} & \textbf{Capstone} & \textbf{\acron} \\
\midrule
Exclusive ownership     & memory sweep  & tree search  & O(1)\\
Virtual memory support   & \xmark & \xmark & \cmark \\
Remote attestation                    & \xmark & \xmark & \cmark \\
In-enclave temporal safety            & \xmark & \cmark & \cmark \\
Source compatibility with Intel \gls{sgx}  & \xmark & \xmark & \cmark \\
Hardware (RTL) implementation         & \cmark & \xmark & \cmark \\
\bottomrule
\end{tabular}
}
\caption{Comparison of capability-based enclave designs.
\cmark~= supported, \xmark~= not supported.}
\label{tab:capcompare}
\end{table}

\paragraph{Capability-based TEEs.}
Most closely related to \acron are CHERI-TrEE~\cite{van2023cheri} and Capstone~\cite{Yu23}. Both extend CHERI toward \gls{tee}-like isolation. Table~\ref{tab:capcompare} compares their key aspects.
CHERI-TrEE~\cite{van2023cheri} gives enclaves unique code/data ownership, local attestation, and secure interrupts, but determines ownership via a full memory sweep at enclave creation.
\acron instead provides an O(1) check and adds remote attestation, in-enclave temporal safety, and support of unmodified SGX applications.
Capstone~\cite{Yu23} provides exclusive access via  \textit{linear capabilities} (move-only, no copies) and \textit{revocation capabilities} for sweep-free reclamation, but
 relies on a custom compiler and hand-written assembly; and its node-tree tracking structure introduces significant performance overhead that \acron avoids. 
Since Capstone lacks an \gls{rtl} implementation, whether maintaining such a node structure is practical in real-world deployments remains unclear. Both \acron and Capstone provide in-enclave temporal safety; \acron additionally provides virtual memory support, remote attestation, and SGX portability.

\ifnotabridged
\section{Discussion  and Future Work}\label{apdx:discussion}

\paragraph{Hue revocation.}
Because a \capcolor is encoded in a fixed-width subfield of the \gls{otype} (\Cref{sec:cam-hw}), the \capcolor namespace is finite: the default 6-bit \texttt{id\_width} allows 63 concurrent enclave instances. Retiring a \capcolor at destruction renders the corresponding \enclavecaps unusable, but does not immediately make the \capcolor re-assignable: stale capabilities bearing that hue may still be held by untrusted software, and re-issuing the \capcolor to a new enclave would let those capabilities alias the new instance's \capvalidtable entry. The \tcb therefore quarantines retired \capcolors and reclaims them in batches. Only once the number of available hues falls below a pre-defined threshold does the \tcb perform a system-wide sweep, walking all tagged memory—and all register state that may hold capabilities—and clearing the tag of every capability whose \gls{otype} encodes a quarantined \capcolor; the reclaimed hues are then returned to the \capcolor allocator. Because the sweep is gated on hue pressure rather than on individual enclave destructions, its cost is amortized across many enclave lifetimes and is fully decoupled from the enclave life cycle: destruction itself remains constant-cost, and no enclave operation blocks on a memory scan. The sweep is also independent of the per-enclave \gls{epvt} revocation sweep of \Cref{sec:temporal-safety}, which is confined to a single enclave's memory and runs in its trusted runtime.

A naïve single-pass sweep is not sound while untrusted software runs concurrently: the host application or \gls{os} holds stale \enclavecaps it may freely copy, so it can read one from a not-yet-swept location and store it into a location the sweep has already passed, surviving the sweep and aliasing a future enclave that receives the recycled hue. Note this is purely a hue-reuse hazard---a retired hue has no active \capvalidtable entry, so the capability cannot be dereferenced.

In our current \acron prototype, the \gls{hmm} shares a single hart (RISC-V hardware thread) with the host \gls{os}, and thus performs the sweep with S-/U-mode execution halted. This approach is trivially sound but requires a pause proportional to physical memory size.
We deem this an acceptable limitation for the prototype, given that availability is already out of scope \Cref{sec:threatmodel} and the sweep is rare by construction, but it makes worst-case latency visible to non-enclave workloads.

We expect multi-core \acron processors to reduce the latency of the system-wide revocation by parallelizing the sweep, similarly to how Cornucopia Reloaded's~\cite{Filardo24} temporal-safety revocation sweep runs in a kernel thread to parallelize its execution relative to userspace code.
In \acron, the \tcb runs in parallel on a hart dedicated for the sweep, while the host \gls{os} is scheduled onto another.
This configuration requires a solution for the above hue-reuse hazard, and we leave evaluating the tradeoffs between the following approaches to future work:

\begin{enumerate}
  \item \textbf{\Enclavecap store barrier.} Add an \\gls{m-mode}-writable \emph{retired-hue mask} (a 63-bit \gls{csr}, or a retiring bit per HAM entry).
  On every capability store from U-/S-mode, compare the stored capability's  \capcolor against the mask and clear the tag on a match. This closes the hazard directly: a retired capability can never be written anywhere, so sweep order is irrelevant and no register filtering is needed. The check is a masked lookup on the hue field, parallel to the existing \capvalidtable \capcolor check.
  \item \textbf{\Enclavecap load barrier.} Apply the same mask check on capability \emph{loads}, clearing the tag in the destination register. Since the only way to move a capability between memory locations is to load it into a register and store it back, a load barrier likewise makes the sweep single-pass and order-independent, and additionally prevents stale capabilities from re-entering the register file at all. It requires one extra step: capabilities already resident in registers when the sweep begins must be filtered, e.g. the \tcb sends a \gls{ipi} to each hart and scrubs retired-\capcolor capabilities from its register file, \glspl{ssa}, and kernel trap frames before the sweep starts. \ifnotabridged Comparing (1) and (2) is largely a question of which path can better absorb the check. \fi
\ifnotabridged
  \item \textbf{Sweep-frontier register.} Maintain an \gls{m-mode} physical-address frontier \gls{csr} marking the swept prefix. Stores from U-/S-mode below the frontier trap to the \tcb, which filters the stored value and completes the store. Only one comparator is added, and the fast path is untouched for the (majority) unswept region; the cost is trap-per-store in the swept region, which grows as the sweep progresses. A \capvalidtable-based variant—temporarily assigning swept regions to the \tcb's own reserved \capcolor reuses existing hardware but is limited by the 64-entry, NAPOT-encoded table.
  \item \textbf{Write-tracked iterative sweep.} Use coarse write tracking (a card table over physical pages, or page-table dirty bits) to re-sweep any page that received stores after being swept, iterating to convergence. Avoids datapath changes but an adversarial \gls{os} can keep dirtying pages indefinitely, so it needs a bounded number of iterations plus a final stop-the-world pass—making it strictly worse than our current stop-the-world sweep unless the tracking granularity is fine enough to make the final pass cheap.
  \item \textbf{Hue epochs.} Pair each hue with a generation counter, held in the \capvalidtable entry and either carried in spare \gls{otype} bits or in an added capability field, so that a recycled hue does not match capabilities from a previous generation. This does not by itself solve the copying hazard, but it enlarges the effective \capcolor namespace and pushes the threshold-triggered sweep further out; with a wide enough epoch, sweeps become rare enough to be comfortably affordable. The trade-off is \gls{otype} pressure: bits taken for epochs come out of the 15-bit color space per enclave (\Cref{sec:cam-hw}).
\fi
\end{enumerate}
\ifabridged
Comparing (1) and (2) is largely a question of which path can better absorb the check.
\fi

\paragraph{Non-contiguous and dynamically resized enclaves.}
A \acron enclave's memory is committed once, at creation, and cannot subsequently change. Because a single \capvalidtable entry describes exactly one \gls{napot} region \Cref{sec:cam-hw}, the kernel backs each enclave with one physically contiguous region whose size if fixed by the \texttt{TEE\_IOCTL\_CREATE} request; the \gls{trts}'s allocator is correspondingly a freestanding allocator over a static arena. An enclave must therefore be provisioned for its worst-case footprint, and over-provisioning is paid for twice: \gls{napot} rounding contributes up to $2\times$ internal fragmentation on top of the requested headroom, and the entire region is zeroed and physically pinned for the enclave's lifetime.
Growth beyond the static enclave boundaries is possible only by \gls{ocall} to into host-managed memory, which lies outside the enclave's \capcolor-protected region and is thus untrusted. This limitation closely mirrors first-generation Intel \gls{sgx}, in which every \gls{epc} page had to be added with \texttt{EADD} before \texttt{EINIT}: the heap was sized at build time from enclave configuration's declared maximum. Intel addresses this in \gls{sgx}2 with \gls{edmm}, which permits pages to be augmented into a running enclave (\texttt{EAUG}) subject to explicit in-enclave acceptance (\texttt{EACCEPT}).
\acron currently provides no analogous facility.

The single-region restriction is an artifact of the \emph{one entry per \capcolor convention}; the \capvalidtable's \texttt{\capcolor} field is per-entry, and nothing in the access check requires \capcolors to be unique across entries. A natural extension for future work is to let \capcolors be associated with a \emph{set of} \gls{napot} regions, added incrementally. We sketch here a candidate design  to support non-contiguous and dynamically resized enclaves, but leave implementation and evaluation for future work.

The \tcb must be permitted to program additional entries carruing and existing \capcolor. The \capcolor-check logic is unchanged: all 64 entries are already evaluated in parallel, and access is granted if /emph{any} matching entry is active with the correct hue. \EInvoke and \EExit must then activate/deactive every entry bearing the target \capcolor rather than a single entry; on a fully-associate table this is a \capcolor-masked broadcast set/clear of the active bit, so domain transitions remain trap-free (\ref{prop:domaintransitions}).

Growth is a privileged operation, since only \gls{m-mode} may program \capvalidtable entures (\Cref{sec:cam-hw}). The enclave requests additional memory from its \gls{trts}; on arena exhaustion it performs an \gls{ocall} asked the host \gls{os} to donate memory; the kernel allocates a \gls{napot} region and issues an \texttt{SBI\_ENCLAVE\_GROW} call over the \texttt{cap\_buffer} protocol. The \tcb then:
\begin{inparaenum}[i)]
  \item rejects any region overlapping a valid \capvalidtable entry, its own reserved region, or a retired \capcolor's region;
  \item zeroes the region under its \gls{m-mode} exception before assigning it to the enclave's \capcolor, so the host \gls{os} cannot chosen bytes into the enclave's heap;
  \item program a new entry with the enclave's \capcolor and sets it valid; and
  \item returns a \enclavecap bound to the new region.
\end{inparaenum}

AS in \gls{sgx} \gls{edmm}, the host \gls{os} must not be able to unilaterally alter an enclave's view of its own address space. The \gls{trts} performs an explicit accept step before the region is joined to the enclave's heap: it validates that the returned \enclavecap contains the enclave's \capcolor, that its bounds fall in a virtual range the \gls{trts} itself requested and that it's disjoint from all existing enclave mappings, and only then links the region into the allocator's free list. A region the enclave has not accepted is never dereferences, so an unsolicited growth is inert.

Following \gls{sgx}2's treatment of augmented pages, grown regions are zero-filled and do not contribute to the enclave measurement, so \texttt{h\_enclave} and the verification procedure of \Cref{sec:ra} are unchanged. The \tcb can additionally record per-\capcolor committed-size high-watermark in the \caphashtable and include it in \texttt{REPORT}, letting a verifier bound the enclave's resource footprint. The \gls{epvt} remains at the top of the primary region and need not grow, since the \gradient-space is fixed by \texttt{pvt\_shift} (\Cref{sec:cam-hw}). Encalve temporal-savety revocation sweeps must now traverse all accepted regions.

The inverse \emph{shrinking} operation follows symmetrically and is worth supporting for long-lived enclaves. On an enclave-initiated release, the \tcb zeroes the region, invalidates its entry, and lets the kernel reclaim the pages; the same \emph{scrub-before-release} ordering as destruction (\Cref{sec:tcb}) is applied to a subset of the enclave's regions. The \tcb must prevent the enclave's primary region from being reclaimed in this manner.

Beyond removing the worst-case provisioning requirements, the design should \emph{reduce} enclave creation latency, the the initial contiguous allocation can start small and grown on demain rather than being sized for the worst case. However, The \capvalidtable is a fixed 64-entry resource shared across up to 63 concurrent hues, so unrestricted growth is not viable. \capvalidtable utilization can be limited by three complementary measures:
\setlist{before=\normalfont,font=\itshape} 
\begin{description}[style=unboxed,leftmargin=0cm]
   \item[Geometric growth:] each grow doubles the committed size.
   \item[Buddy coalescing:] when a new region is the aligned buddy of an existing one, the HMM merges them into a single entry, so the steady-state entry count per enclave stays small.
   \item[Region table with \capvalidtable as a cache:] for the general case, maintain a per-hue region table in \tcb-private memory and treat the \capvalidtable as a fully-associative cache over it, filled on miss by a hardware walker or an \gls{m-mode} fill trap. This is conceptually the same approach as in \gls{cove} with its hardware-walked \gls{mtt}~\cite{Sahita23} (see below), decoupling the maximum number of regions from the table size at the cost of handling misses.
\end{description}

\paragraph{Susceptibility of SGX Enclave code to memory-safety issues.}
A key advantage of \acron is that enclave memory safety is enforced by default for all code, including the off-the-shelf SGX applications. Running these atop \acron surfaces latent memory-safety bugs that SGX's lack of hardware bounds enforcement allows to go silently unnoticed. 

For example, for the examples used in our performance evaluation (in~\Cref{sec:eval}), \texttt{CryptoEnclave} passes miscounted buffer lengths into its cryptographic routines: NUL-inclusive lengths, an unstripped \texttt{strlen+1} key, and its AES decryption loop reads a full 4~KiB block from a ciphertext buffer sized to a smaller per-chunk allocation, overreading the heap. Similarly, \texttt{SGX\_SQLite}'s \texttt{btree} integrity checker uses an unbounded write index into a fixed-size coverage buffer, causing \texttt{btreeHeapInsert()} to overflow whenever a page's cell and freeblock count exceeded \texttt{pageSize/4}.
This underscores that the lack of inherent memory-safety measures in the \gls{tee} programming model manifests as silent bugs in real-world applications, whereas \acron's capability foundation catches these violations.

Our source code changes to address these and other compatibility issues add or modify the following number of lines for each application: 16 for \texttt{CryptoEnclave}, 25 for \texttt{sgx-kmeans}, and 16 for \texttt{SGX\_SQLite}. This excludes minor changes, such as removal of SGX EID's from function signatures, adding \texttt{extern ``C''} labels in function declarations for compiling with our runtimes, and renaming of functions to avoid conflicts with C libraries (e.g., \texttt{printf}).  

\paragraph{Speculative-execution and side-channel threat surface.}
\PRISM does not attempt to systematically mitigate microarchitectural side-channels, and we treat transient-execution and timing leakage as orthogonal to the isolation guarantees established in \Cref{sec:highlevel}.
This scoping decision warrants justification, particularly in comparison to \gls{sgx}.

A substantial share of the attacks that have eroded \gls{sgx}'s security guarantees are not failures of its isolation logic but artifacts of the commodity x86\_64 microarchitecture into which SGX is embedded.
Transient-execution attacks such as Foreshadow~\cite{VanBulck18}, LVI~\cite{VanBulck20}, and the broader Spectre/Meltdown family~\cite{Kocher19,Lipp18,Horn18,Wikner25} exploit speculative and out-of-order machinery that is shared with the rest of the core; \gls{sgx} enclaves inherit this attack surface rather than introduce it.

\gls{sgx} is thus perennially exposed to a class of vulnerabilities that are entangled with decades of accreted x86 microarchitectural complexity and are difficult to retrofit against without invasive changes or steep performance penalties.
\PRISM, by contrast, is built on an open CHERI-RISC-V substrate whose speculative behavior is comparatively constrained and, crucially, tractable to reason about: recent work on architectural contracts for safe speculation in CHERI~\cite{Fuchs23,Fuchs24} show that principled, hardware-enforced mitigation of transient leakage can be integrated into this platform rather than bolted on.

We emphasize that this is a difference in the \emph{tractability and extent of the threat surface}, not a claim of immunity; an out-of-order core such as CHERI-Toooba retains a transient-execution surface, but the absence of \gls{sgx}'s legacy microarchitecture means \PRISM does not begin from the same disadvantaged position, and the mitigations above can compose with \PRISM.

We argue that pushing complete side-channel resistance into the hardware architecture is neither practical nor, in general, desirable.
Enclave workloads differ widely in their side-channel objectives: a workload processing only public inputs may reasonably tolerate data-dependent timing, whereas one manipulating long-term secrets demands constant-time execution and data-oblivious memory access.
A one-size-fits-all architectural guarantee would either under-serve the latter or impose its cost on the former.
This tension is sharpest for cryptographic code, where the appropriate defenses are algorithm- and even implementation-specific.
For  example, the side-channel-resistant realization of post-quantum schemes, in particular, remains an open research problem; the leakage surface of lattice-based and other post-quantum constructions is still being mapped, and constant-time or masked implementations continue to be discovered vulnerable.

Fixing a side-channel countermeasure in silicon under these conditions risks entrenching a defense that later proves incomplete. \PRISM therefore deliberately factors side-channel resistance out of its isolation primitives and leaves it to enclave software and to composable mechanisms, e.g., data-oblivious computation over blinded capabilities as in BLACKOUT~\cite{ElAtali25}, that can be applied selectively according to each enclave's threat model.
This separation of concerns keeps \PRISM's \gls{tcb} minimal while leaving the door open to stronger, use-case-appropriate guarantees where they are warranted.

\glsreset{cove}
\glsreset{mtt}
\paragraph{\PRISM and RISC-V \glspl{cvm}.}
\PRISM and RISC-V's emerging confidential-computing stack occupy complementary points in the isolation design space: the RISC-V \gls{cove}~\cite{Sahita23} provides confidential computing through a \gls{tsm} operating in M-mode to mediate transitions between confidential and non-confidential contexts, with the TSM acting as a trusted intermediary between the hypervisor and its \glspl{tvm}.
Memory ownership in \gls{cove} is enforced by a hardware-walked \gls{mtt} with per-domain access permissions, so that once physical memory is converted to confidential it is reachable only by the confidential supervisor domain.
A \gls{tvm} is a coarse-grained, whole-VM abstraction: the unit of protection is a guest \gls{os} and its address space, and the \gls{tcb} inside the \gls{tvm} is correspondingly large.

\PRISM and \gls{cove} differ in three aspects which makes them complementary.
First, in \emph{granularity}: a \gls{tvm} isolates an entire \gls{vm} from the host, whereas a \PRISM enclave isolates a sub-region of a single user process (at object-level capability granularity), without a guest \gls{os}.
Second, in \emph{isolation mechanism}: \gls{cove} partitions physical memory between supervisor domains through the \gls{mtt}, conceptually analogous to Intel \gls{tdx}'s Secure-EPT or Arm \gls{cca}'s Granule Protection Table, and enforces protection at the granularity of pages assigned to a domain; \PRISM instead binds physical ownership to unforgeable, M-mode-rooted hues carried in the capabilities themselves, and additionally inherits \gls{cheri}'s intra-enclave spatial and temporal memory safety, a guarantee entirely absent from the whole-\gls{vm} model, whose in-TVM memory-safety posture is no better than a conventional guest.
Third, in \emph{threat surface}: CoVE assumes a trusted \gls{os} and runtime inside the \gls{tvm}, whereas \PRISM's threat model treats even the host \gls{os} as adversarial and shrinks the trusted computing base to a minimal M-mode monitor and the enclave code itself.

A compelling direction for future work is composing these two mechanism by nesting \PRISM enclaves \emph{within} a \gls{cove} \gls{tvm}.
A nested arrangement would yield defense-in-depth and let a \gls{cvm} defend against a malicious hypervisor and host, while \PRISM enclaves inside it defend sensitive components against the \gls{tvm}'s own (potentially compromised) guest \gls{os} and against memory-safety exploits.

Realizing this composition raises open questions we leave to future work. Chiefly, reconciling the \gls{mtt}'s and \gls{cam}'s views of physical memory ownership, extending \PRISM's attestation chain to the \gls{cove} layered-attestation architecture so that a verifier can reason about the enclave, the enclosing \gls{tvm}, and the platform in a single evidence chain, and defining the interrupt- and world-switch interactions when both the \gls{tsm} and the \gls{hmm} contend for M-mode traps.

\fi

\ifabridged
\section{Future Work}\label{sec:future-work}

\PRISM provides isolated enclave execution, spatial and temporal memory safety, and remote attestation as its key security properties. A remaining attack surface is through transient-execution and microarchitectural side channels, a common concern in many \gls{tee} architectures. Future work can incorporate mechanisms like BLACKOUT~\cite{ElAtali25} to enable data-oblivious computation via blinded capabilities.
Another future work direction is exploring how to integrate \PRISM with \gls{cvm} extensions. Due to their differences in isolation granularity and mechanisms along with differences in threat surface, a nested design of \PRISM enclaves within a \gls{cvm} is an intriguing direction for future work.
We discuss further details regarding side-channel threat surface and adapting \PRISM to RISC-V \glspl{cvm} in Appendix~\ref{apdx:discussion}.
\else
\section{Conclusion}\label{sec:conclusion}

We introduced \enclavecaps, a \gls{cheri} extension that binds an enclave's capabilities to an unforgeable, M-mode-rooted \capcolor enforcing exclusive access to physical memory, and \PRISM, a \gls{tee} architecture realizing them on CHERI-RISC-V.

By carrying ownership in the capability itself and checking it against a hardware hue-addressed table after translation, \PRISM resolves the design challenges that have constrained prior capability-based \glspl{tee}: 
\begin{inparaenum}[1)]
  \item efficient establishment of exclusive-ownership of a set of capabilities without memory sweeps or revocation trees at enclave initialization time, 
  \item enforcement of physical-memory exclusivity even against a malicious OS that controls address translation, performs domain transitions through unprivileged,
  \item trap-free enclave entry, 
  \item and supports remote attestation rooted in a minimal M-mode monitor.
\end{inparaenum}

We implemented \PRISM on both a CHERI-RISC-V QEMU emulator and a CHERI-Toooba \gls{fpga} softcore and equipped it with support in CheriBSD and an SGX-compatibility layer that runs unmodified enclaves, and evaluated it across three real-world SGX applications with modest hardware cost ($\approx16\%$ logic) and low run-time overhead ($\approx6-15$\%).

A symbolic Tamarin analysis establishes the security of its attestation protocol. 
Together these results show that strong userspace enclaves can be built as a lightweight, capability-native extension.
\fi

\ifnotanonymous
\section*{Acknowledgments}

We thank our colleagues at Ericsson: Daniel Migault and Santeri Paavolainen for their feedback on this manuscript.
We are deeply grateful to Hossam ElAtali for his feedback on the design and the numerous discussion we had over the course of the work.
This work is supported in part by the Wallenberg Visiting
Professor Program and the Natural Sciences and Engineering
Research Council of Canada (grant number RGPIN-2026-
04826).
\fi

\bibliographystyle{IEEEtran}
\bibliography{references, main}

\ifnotarxiv
\appendices
\else
\appendix
\fi

\section{Additional Performance Evaluation}\label{apdx:prism-picasso-sgx}

We extend the evaluation from Section~\ref{sec:eval} to also compare the additional end-to-end run-time cost of \acron (enclave creation, temporally safe execution, enclave destruction) to running the same application atop \texttt{PICASSO} (which enables temporally safe execution). To obtain the non-enclave baseline application, we modify the source code to remove the enclave creation and destruction operations, and adjust the C-file header inclusions so that all code paths remain reachable. 

\begin{figure*}[t]
    \begin{subfigure}{\textwidth}
  \centering
  \resizebox{\textwidth}{!}{
  \begin{tikzpicture}
    \begin{groupplot}[
      group style={
        group size=10 by 1,
        horizontal sep=20pt,
        x descriptions at=edge bottom,
      },
      width=\appSubPlotWidth,
      height=\appSubPlotHeight,
      ybar=1pt,
      ymin=0,
      symbolic x coords={T},
      xtick={T},
      yticklabel style={font=\normalsize},
      xticklabel style={font=\normalsize},
      scale only axis,
      enlarge x limits=0.2,
      cycle list={
        {fill=barwhite, draw=black, thick},
        {fill=barlightgray, draw=black, thick},
        {fill=bardarkgray, draw=black, thick},
        {fill=barblack, draw=black, thick},
      }
    ]
    
    \nextgroupplot[ylabel={Milliseconds (ms)}, xticklabels={SHA256-T}]
    \addplot coordinates {(T,0.164)}; \addplot coordinates {(T,16.84)}; \addplot coordinates {(T,0.02)};;

    \nextgroupplot[xticklabels={SHA256-F}]
    \addplot coordinates {(T,66.78)}; \addplot coordinates {(T,84.22)}; \addplot coordinates {(T,67.42)};

    \nextgroupplot[xticklabels={HMAC256-T-I}]
    \addplot coordinates {(T,0.178)}; \addplot coordinates {(T,16.82)}; \addplot coordinates {(T,0.02)};

    \nextgroupplot[xticklabels={HMAC256-T-R}]
    \addplot coordinates {(T,0.1797)}; \addplot coordinates {(T,16.8359	)}; \addplot coordinates {(T,0.0234)};

    \nextgroupplot[xticklabels={HMAC256-F-I}]
    \addplot coordinates {(T,66.7813)}; \addplot coordinates {(T,84.2031)}; \addplot coordinates {(T,67.3828)};

    \nextgroupplot[xticklabels={HMAC256-F-R}]
    \addplot coordinates {(T,66.7500)}; \addplot coordinates {(T,84.1953)}; \addplot coordinates {(T,67.3906)};

    \nextgroupplot[xticklabels={AES-ECB-F-R}]
    \addplot coordinates {(T,0.1875)}; \addplot coordinates {(T,16.8125)}; \addplot coordinates {(T,0.0234)};

    \nextgroupplot[xticklabels={AES-ECB-T-R}]
    \addplot coordinates {(T,6.7656)}; \addplot coordinates {(T,23.9219)}; \addplot coordinates {(T,7.1172)};

    \nextgroupplot[xticklabels={AES-CBC-F-R}]
    \addplot coordinates {(T,0.1875)}; \addplot coordinates {(T,16.8516)}; \addplot coordinates {(T,0.0234)};

    \nextgroupplot[xticklabels={AES-CBC-T-R}]
    \addplot coordinates {(T,8.0313)}; \addplot coordinates {(T,25.2188)}; \addplot coordinates {(T,8.4063)};

    \end{groupplot}
    \end{tikzpicture}
  }
  \caption{Run-time of \texttt{CryptoEnclave} test cases in milliseconds (ms). Test cases vary based on the particular cryptographic operation,  command line text (T) or file (F) input, and a random (R) or input (I) key.}
  \label{fig:crypto-run-time}
\end{subfigure}
    \begin{subfigure}{\textwidth}
  \centering
  \resizebox{\textwidth}{!}{
  \begin{tikzpicture}
    \begin{groupplot}[
      group style={
        group size=10 by 1,
        horizontal sep=30pt,
        x descriptions at=edge bottom,
      },
      width=\appSubPlotWidth,
      height=\appSubPlotHeight,
      ybar=1pt,
      ymin=0,
      symbolic x coords={T},
      xtick={T},
      yticklabel style={font=\normalsize},
      xticklabel style={font=\normalsize},
      scale only axis,
      enlarge x limits=0.2,
      cycle list={
        {fill=barwhite, draw=black, thick},
        {fill=barlightgray, draw=black, thick},
        {fill=bardarkgray, draw=black, thick},
        {fill=barblack, draw=black, thick},
      }
    ]
    \nextgroupplot[ylabel=Seconds(s), xticklabels={N=100K}]
    \addplot coordinates {(T,180.78)}; \addplot coordinates {(T,196.22)}; \addplot coordinates {(T,179.51)};;
    \nextgroupplot[xticklabels={N=200K}]
    \addplot coordinates {(T,339.80)}; \addplot coordinates {(T,357.25)}; \addplot coordinates {(T,340.52)};
    \nextgroupplot[xticklabels={N=300K}]
    \addplot coordinates {(T,1351.00)}; \addplot coordinates {(T,1378.27)}; \addplot coordinates {(T,1361.59)};
    \nextgroupplot[xticklabels={N=400K}]
    \addplot coordinates {(T,1319.00)}; \addplot coordinates {(T,1338.89)}; \addplot coordinates {(T,1322.18)};
    \nextgroupplot[xticklabels={N=500K}]
    \addplot coordinates {(T,1678.11)}; \addplot coordinates {(T,1721.41)}; \addplot coordinates {(T,1704.68)};
    \nextgroupplot[xticklabels={N=600K}]
    \addplot coordinates {(T,1685.96)}; \addplot coordinates {(T,1233.27)}; \addplot coordinates {(T,1216.56)};
    \nextgroupplot[xticklabels={N=700K}]
    \addplot coordinates {(T,1513.68)}; \addplot coordinates {(T,1525.45)}; \addplot coordinates {(T,1508.72)};
    \nextgroupplot[xticklabels={N=800K}]
    \addplot coordinates {(T,1878.26)}; \addplot coordinates {(T,1992.55)}; \addplot coordinates {(T,1975.82)};
    \nextgroupplot[xticklabels={N=900K}]
    \addplot coordinates {(T,2825.21)}; \addplot coordinates {(T,2825.52)}; \addplot coordinates {(T,2808.81)};
    \nextgroupplot[xticklabels={N=1M}]
    \addplot coordinates {(T,2504.89)}; \addplot coordinates {(T,2837.51)}; \addplot coordinates {(T,2820.83)};
    \end{groupplot}
    \end{tikzpicture}
  }
  \vspace{-1.5em}
  \caption{Run-time of \texttt{sgx-kmeans} test cases in milliseconds. Test cases use randomly generated input files of sizes 100K to 1M varying by 100K. The average time for K from 2 to 8 is reported.}
  \label{fig:kmeans-run-time}
  \vspace{-1.5em}
\end{subfigure}
    \begin{subfigure}{\textwidth}
  \centering
  \resizebox{\textwidth}{!}{
  \begin{tikzpicture}
    \begin{groupplot}[
      group style={
        group size=10 by 1,
        horizontal sep=30pt,
        x descriptions at=edge bottom,
      },
      width=\appSubPlotWidth,
      height=\appSubPlotHeight,
      ybar=1pt,
      ymin=0,
      symbolic x coords={T},
      xtick={T},
      yticklabel style={font=\normalsize},
      xticklabel style={font=\normalsize},
      scale only axis,
      enlarge x limits=0.2,
      cycle list={
        {fill=barwhite, draw=black, thick},
        {fill=barlightgray, draw=black, thick},
        {fill=bardarkgray, draw=black, thick},
        {fill=barblack, draw=black, thick},
      }
    ]

    \nextgroupplot[ylabel={Milliseconds(s)}, xticklabels={\texttt{select1}}]
    \addplot coordinates {(T,98.05)}; \addplot coordinates {(T,111.34)}; \addplot coordinates {(T,92.84)};

    \nextgroupplot[xticklabels={\texttt{select2}}]
    \addplot coordinates {(T,87.98)}; \addplot coordinates {(T,101.34)}; \addplot coordinates {(T,82.66)};

    \nextgroupplot[xticklabels={\texttt{select3}}]
    \addplot coordinates {(T,254.54)}; \addplot coordinates {(T,257.62)}; \addplot coordinates {(T,239.08)};

    \nextgroupplot[xticklabels={\texttt{ib\_1\_0}}]
    \addplot coordinates {(T,202.40)}; \addplot coordinates {(T,233.59)}; \addplot coordinates {(T,214.38)};

    \nextgroupplot[xticklabels={\texttt{ib\_10\_0}}]
    \addplot coordinates {(T,293.30)}; \addplot coordinates {(T,330.20)}; \addplot coordinates {(T,310.96)};

    \nextgroupplot[xticklabels={\texttt{ib\_100\_0}}]
    \addplot coordinates {(T,939.70)}; \addplot coordinates {(T,1007.50)}; \addplot coordinates {(T,988.15)};

    \nextgroupplot[xticklabels={\texttt{ib\_1000\_0}}]
    \addplot coordinates {(T,1793.41)}; \addplot coordinates {(T,1908.45)}; \addplot coordinates {(T,1889.13)};

    \end{groupplot}
    \node[anchor=west, inner sep=26pt] at (current axis.east) {
        \begin{tabular}{{|l|}}
          \hline
          \textbf{Legend:}\\
          \tikz{\fill[barwhite,draw=black,thick] (0,0) rectangle (6pt,6pt);} \PICASSO Total\\
          \tikz{\fill[barlightgray,draw=black,thick] (0,0) rectangle (6pt,6pt);} \acron Total\\
          \tikz{\fill[bardarkgray,draw=black,thick] (0,0) rectangle (6pt,6pt);} \acron Exec.\\
          \hline
        \end{tabular}
      };
    \end{tikzpicture}
  }
  \caption{Run-time of \texttt{SGX\_SQLite} shown in seconds for \texttt{select} and \texttt{index\_between} (\texttt{ib}) examples from \texttt{sqllogictest}~\cite{sqllogictest} test suite.}
  \vspace{-1.5em}
  \label{fig:sqlite-run-time}
\end{subfigure}
    \caption{Comparison between execution of enclave applications atop \texttt{PICASSO} vs. \acron. Three times are presented: \texttt{PICASSO} total end-to-end time, \acron total end-to-end time, and \acron execution time (excluding enclave create and destroy times).}
    \label{fig:app-run-time}
\end{figure*}
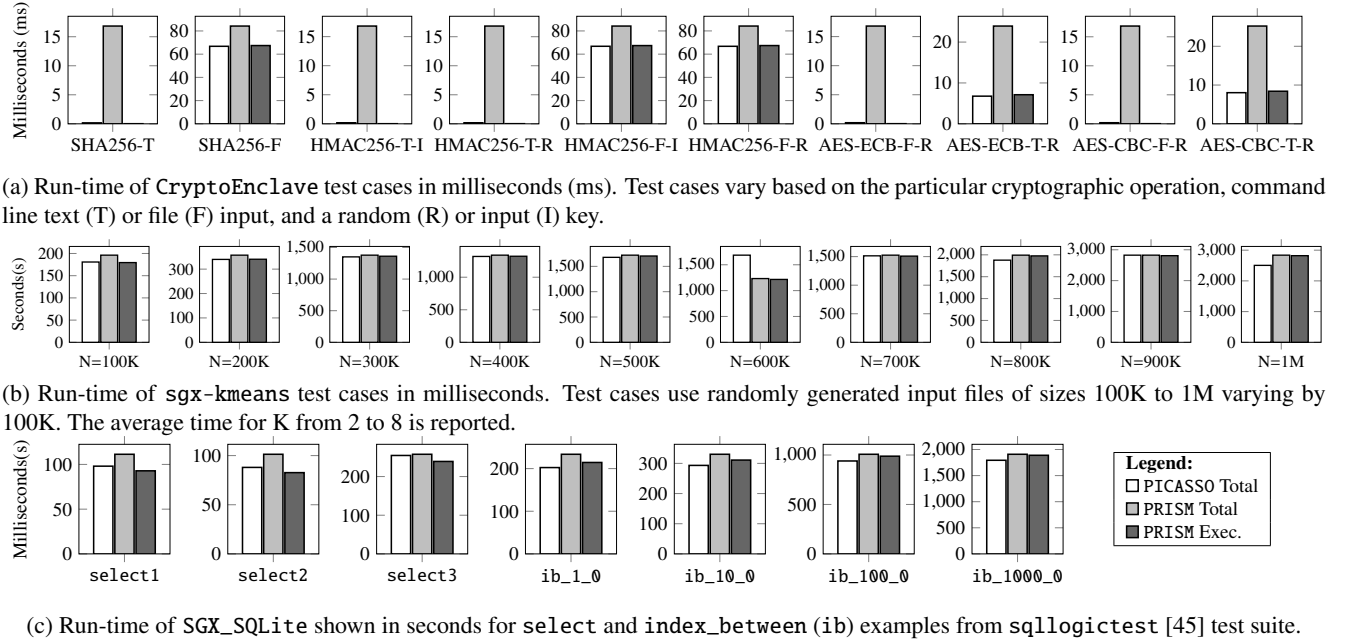

Figure~\ref{fig:app-run-time} compares run time of the example SGX applications on \acron vs. \texttt{PICASSO}. For \acron's run time, we report two measurements:
\begin{itemize}[nosep]
    \item end-to-end run time: enclave creation, temporally safe execution, and enclave destruction (denoted ``\acron Total'');
    \item execution time: the temporally safe and isolated execution time alone (denoted ``\acron Exec.'').
\end{itemize}
For \texttt{PICASSO}, we report only execution time, since it does not support enclave-specific operations by default. 

For \texttt{CryptoEnclave}, shown in Figure~\ref{fig:crypto-run-time}, the performance overhead varies greatly depending on the particular configuration. When the cryptographic operation takes input text via command line (denoted with ``T''), execution time itself is negligible. This demonstrates that for simpler executions, the majority of \acron overhead stems from life-cycle operations (e.g., enclave creation and destruction). For the variants that take input files (denoted with ``F''), the execution time is more significant, reflected in the smaller gap between \acron's end-to-end time and execution time. For the file-taking operations, \acron introduces 0.95\%--4.46\% overhead onto the execution time, whereas enclave life-cycle operations raise end-to-end run-time overhead to 20.7\%--71.7\%. 

For \texttt{SGX\_SQLite} test cases, shown in Figure~\ref{fig:sqlite-run-time}, \acron adds 4.9\%--5.7\% execution overhead to \texttt{select} tests, while reducing execution overhead for \texttt{ib} tests by 5.6\%--6.4\%. 
Including the additional costs for enclave creation and destruction, the end-to-end overhead accounts for a 1.19\%-13.3\% increase. 

A similar pattern is observed in \texttt{sgx-kmeans}, shown in Figure~\ref{fig:kmeans-run-time}: for pure execution time, \acron reduces overhead by 0.33\%--0.70\% for $N=$ 100K, 700K, and 900K, while adding overhead of up to 11.7\% for other input sizes. This minimal variation shows that for these test cases, execution overhead is essentially negligible. The end-to-end overhead of \acron for \texttt{sgx-kmeans} test cases ranges from 0.01\%--11.7\%, heavily influenced by the marshaling of extracted data points via ECALL from the host application into the enclave.

\ifnotarxiv

\fi
\section{Tamarin Model}\label{apdx:tamarin}

We model \acron's measurement and remote attestation process in Tamarin. We model the mechanisms and protocol steps from Section~\ref{sec:ra} as a multiset of rules, and each sub-security property of \ref{prop:attestation} from Section~\ref{sec:security} using lemmas. We use the \texttt{hashing}, \texttt{signing}, and \texttt{natural-numbers} Tamarin built-ins, providing modeling of a collision-resistant hash function, digital signature functions, and natural-number arithmetic, respectively. The rule-based model is shown in Figure~\ref{fig:tamarin-rules}, and the lemmas are shown in Figure~\ref{fig:tamarin-lemmas}. 

\begin{figure*}[t]
\begin{subfigure}[b]{0.5\textwidth}
\centering
\begin{lstlisting}[style=tamarinstyle]
|\ruleheader{Install HMM}\label{tamarin:rule:install_tcb}|
rule Install_HMM:
  let
     h_exp_hmm = h(~S_HMM)
  in
  [Fr(~sk_prv), Fr(~S_HMM)]
  --[HMM_Install(~sk_prv, h_exp_hmm), SecretKey(~sk_prv)]->
  [!System_Init(~sk_prv, ~S_HMM),!Pk(pk(~sk_prv)),
     Out(pk(~sk_prv)), !HueTable(

|\ruleheader{Create Encalve}\label{tamarin:rule:create_encalve}|
rule Create_enclave:
  [Fr(~S_encl), !HueTable(
  --[Encl_Install(~S_encl,
  [!HueTable(

|\ruleheader{Attestation Request}\label{tamarin:rule:att_request}|
rule Attestation_Request:
  [In(<
  --[ReceivedRequest(
  [Get_Report(

|\ruleheader{Generate Report}\label{tamarin:rule:gen_report}|
rule Generate_Report:
  let
    h_hmm = h(S_HMM)
    sig = sign(<
  in
  [!System_Init(k, S_HMM), Get_Report(
  --[ Signed(
  [ Out(<<

|\ruleheader{Verify Signature}\label{tamarin:rule:verify}|
rule Verify_Signature:
  [In(<report_data, sig>), !Pk(pk_prv)]
  --[Verified(report_data, verify(sig, report_data, pk_prv), pk_prv)]->
  []
\end{lstlisting}
\end{subfigure}
~
\begin{subfigure}[b]{0.5\textwidth}
\centering
\begin{lstlisting}[style=tamarinstyle]
|\ruleheader{Install HMM}\label{tamarin:rule:install_tcb}|
rule Install_HMM:
  let
     h_exp_hmm = h(~S_HMM)
  in
  [Fr(~sk_prv), Fr(~S_HMM)]
  --[HMM_Install(~sk_prv, h_exp_hmm), SecretKey(~sk_prv)]->
  [!System_Init(~sk_prv, ~S_HMM),!Pk(pk(~sk_prv)),
     Out(pk(~sk_prv)), !HueTable(

|\ruleheader{Create Encalve}\label{tamarin:rule:create_encalve}|
rule Create_enclave:
  [Fr(~S_encl), !HueTable(
  --[Encl_Install(~S_encl,
  [!HueTable(

|\ruleheader{Attestation Request}\label{tamarin:rule:att_request}|
rule Attestation_Request:
  [In(<
  --[ReceivedRequest(
  [Get_Report(

|\ruleheader{Generate Report}\label{tamarin:rule:gen_report}|
rule Generate_Report:
  let
    h_hmm = h(S_HMM)
    sig = sign(<
  in
  [!System_Init(k, S_HMM), Get_Report(
  --[ Signed(
  [ Out(<<

|\ruleheader{Verify Signature}\label{tamarin:rule:verify}|
rule Verify_Signature:
  [In(<report_data, sig>), !Pk(pk_prv)]
  --[Verified(report_data, verify(sig, report_data, pk_prv), pk_prv)]->
  []
\end{lstlisting}
\end{subfigure}
\caption{Rule-based model of \acron measurement and attestation mechanism.}\label{fig:tamarin-rules}
\end{figure*}

\subsection{Rule-based Model.}

Each rule follows the following structure: (1) has a set of precondition facts that are required to be true before the rule can be evaluated, (2) a set of action facts in the middle which are recorded when the precondition facts exist, and (3)  a set of postcondition facts that are produced to complete the evaluation of the rule.  

Rule~\ref{tamarin:rule:install_tcb} models the initial step for installing the \gls{hmm} into the system. The precondition shows that two \emph{fresh} values (denoted with \texttt{\textcolor{blue}{Fr}}) must be generated: one for the \prv's secret key (\texttt{sk\_prv}) and another for the \gls{hmm} software itself (\texttt{S\_HMM}). When this precondition is met, two action facts are recorded. First, \texttt{HMM\_Install} is recorded to show that the \gls{hmm} itself was indeed installed with \texttt{S\_HMM} and \texttt{SecretKey} to show that  the \gls{hmm} exists with \texttt{sk\_prv}.
Four facts comprise the postcondition of this rule, which can then be used as preconditions for other rules:
\begin{itemize}
    \item \texttt{System\_Init}, used to signal that the system is initialized with \texttt{S\_HMM} and \texttt{sk\_prv}; 
    \item \texttt{Pk}, signaling that a public key symbol corresponding to \texttt{sk\_prv} exists;
    \item \texttt{\textcolor{blue}{Out}}, a built-in fact supplied by Tamarin to represent when a symbol is output to the network. In this case, the public key of \texttt{sk\_prv} is broadcast into the network;
    \item \texttt{HueTable} to demonstrate that an empty but initialized \caphashtable was generated. In Tamarin, natural-numbers start at 1, so the counter representing no entries in the HueTable uses \texttt{\%1}.
\end{itemize}

Rule~\ref{tamarin:rule:create_encalve} models enclave creation. The precondition for this rule is that a fresh enclave software (\texttt{S\_encl}) exists, and a \texttt{HueTable} fact exists with an arbitrary \texttt{last\_hue} assigned. This means that the creation of an enclave cannot occur unless a \texttt{HueTable} was initialized.
When these preconditions are met, this rule logs one action fact used later for verification of lemmas: \texttt{Encl\_install}, recording the software and the \capcolor that will be assigned to it, by incrementing \texttt{last\_hue}. The post condition is that another \texttt{HueTable} fact is generated for the created enclave, logging the \capcolor and the hash of the enclave code (\texttt{\textcolor{blue}{h}(S\_encl)}). 

Rule~\ref{tamarin:rule:att_request} corresponds to \prv receiving an attestation request from \vrf. A precondition is that a \texttt{hue} and \texttt{chal} value must be received together over the network. When this occurs, the action fact \texttt{ReceivedRequest} is created to log them. A postcondition of this rule is the generation of the fact \texttt{Get\_Report}, symbolizing that the host has requested a report upon receipt of the request.  

Rule~\ref{tamarin:rule:gen_report} models the generation of the report by the \gls{hmm}. The precondition of this rule is that (1) the \acron system was initialized, (2) a request was received, and (3) an enclave with the requested \texttt{hue} was created. These three are represented by the \texttt{System\_Init}, \texttt{Get\_Report}, and \texttt{HueTable} precondition facts, respectively.
When these preconditions are met, a report can be generated by producing a signature over the contents: \texttt{\%hue, chal, h\_hmm, h\_encl}.
This is logged with the action fact \texttt{Signed}, which records the contents that the signature \texttt{sig} is computed over. The postcondition of this rule is that the packed report is sent \texttt{\textcolor{blue}{Out}} into an untrusted channel. 

Finally, Rule~\ref{tamarin:rule:verify} has two preconditions: it receives some packed input structure via \texttt{\textcolor{blue}{In}} containing some arbitrary \texttt{report\_data} and \texttt{sig}, and the public key of a prover exists (e.g., \texttt{Pk} fact from Rule~\ref{tamarin:rule:install_tcb}). When this precondition is met, an action fact \texttt{Verified} is logged to acknowledge the verification attempt. This records the \texttt{report\_data} that is being verified, the public key used for verification \texttt{pk\_prv}, and the verification result, using the built-in function \texttt{verify}. This rule has no postcondition, as it models the protocol's last step.

\subsection{Security Sub-properties.}

Figure~\ref{fig:tamarin-lemmas} shows the Tamarin lemmas modeling the sub-security properties of \ref{prop:attestation}.
Lemma~\ref{tamarin:lemma:sig_check} is used to check report authenticity. It specifically outlines: a report verification result returning as true implies that \acron \gls{hmm} produced that report. Lemma~\ref{tamarin:lemma:hmm_check} is used to check \gls{hmm} integrity.
This specifies that if a report verification including an \gls{hmm} hash of \texttt{h\_hmm} returns as true, there must have been an \gls{hmm} software installed at a prior point with a hash \texttt{h\_hmm\_exp}, and \texttt{h\_hmm\_exp} must equal \texttt{h\_hmm}. Lemma~\ref{tamarin:lemma:encl_check} checks enclave integrity and follows a similar structure.
It checks that if a report verification returning true contains enclave hash \texttt{h\_encl} and \texttt{hue}, there must exist an enclave installation by the \gls{hmm} that has the same \capcolor-hash pair.

Lemma~\ref{tamarin:lemma:key_secret} models that a secret key is neither exposed over an \adv-observable channel nor is derivable from data observed on these channels.
This is expressed by checking: if there exists a \texttt{SecretKey(k)}, there does not exist any time when \adv can learn \texttt{k}.
This is modeled with Tamarin built-in rule \texttt{K} to model \adv's knowledge of a symbol.

Finally, Lemma~\ref{tamarin:lemma:freshness} ensures that any response that passes verification uses a fresh attestation challenge, showing that responses must be produced per-request.
This is modeled in a similar manner to Lemmas~\ref{tamarin:lemma:hmm_check} and~\ref{tamarin:lemma:encl_check}: a verification result returning true using an attestation challenge \texttt{chal} implies there must exist a prior attestation request containing \vrf's challenge (\texttt{vrf\_chal}), and \texttt{vrf\_chal} equals \texttt{chal}.

These lemmas all pass verification with the results described in Section~\ref{sec:security}.

\begin{figure}[t]
\begin{lstlisting}[style=tamarinstyle]
|\lemmaheader{Signature Check}\label{tamarin:lemma:sig_check}|
lemma signature_check:
 "All hue chal h_hmm h_encl pk_prv #j.
  Verified(<hue, chal, h_hmm, h_encl>, true, pk_prv) @ j
  & not (hue =
    ==> (
        Ex  #i.
        Signed(hue, chal, h_hmm, h_encl) @ i
        & i < j
    )"

|\lemmaheader{HMM Code Check}\label{tamarin:lemma:hmm_check}|
lemma hmm_code_check:
 "All hue chal h_hmm h_encl pk_prv #j.
  Verified(<hue, chal, h_hmm, h_encl>, true, pk_prv) @ j
    ==> (
        Ex #i sk_prv h_hmm_exp.
        HMM_Install(sk_prv, h_hmm_exp) @ i
        & (h_hmm = h_hmm_exp)
        & i < j
    )"

|\lemmaheader{Enclave Code Check}\label{tamarin:lemma:encl_check}|
lemma encl_code_check:
 "All hue chal h_hmm h_encl pk_prv #j.
  Verified(<hue, chal, h_hmm, h_encl>, true, pk_prv) @ j
  & not (hue =
    ==> (
        Ex s_encl_exp #i.
        Encl_Install(s_encl_exp, hue) @ i
        & (h_encl = h(s_encl_exp))
        & i < j
    )"

|\lemmaheader{Key Secrecy}\label{tamarin:lemma:key_secret}|
lemma key_secrecy:
  "All k #i. SecretKey(k) @ #i ==> not (Ex #j. K(k) @ #j)"

|\lemmaheader{Freshness}\label{tamarin:lemma:freshness}|
lemma freshness:
 "All hue chal h_hmm h_encl pk_prv #j.
  Verified(<hue, chal, h_hmm, h_encl>, true, pk_prv) @ j
    ==>
        Ex vrf_chal #i.
        ReceivedRequest(hue, vrf_chal) @ i
        & (chal = vrf_chal)
        & i < j
    "
\end{lstlisting}
\caption{Tamarin Lemmas, each corresponding to a sub-security property, which all together provide \ref{prop:attestation}.}
\label{fig:tamarin-lemmas}
\end{figure}

\end{document}